\documentclass[sigconf]{acmart}
\pdfoutput=1

\def\BibTeX{{\rm B\kern-.05em{\sc i\kern-.025em b}\kern-.08emT\kern-.1667em\lower.7ex\hbox{E}\kern-.125emX}}

\acmConference[CCS’24]{Conference on Computer and Communications Security}{Oct 14--18}{Salt Lake City, USA}

\AtBeginDocument{%
  \providecommand\BibTeX{{%
    Bib\TeX}}}

\usepackage{nicefrac}
\usepackage{siunitx}
\usepackage{array,framed}
\usepackage{booktabs}
\usepackage{
  color,
  float,
  epsfig,
  wrapfig,
  graphics,
  graphicx,
  subcaption
}
\usepackage{textcomp}
\usepackage{setspace}
\usepackage{amsmath}
\usepackage{amsfonts}
\usepackage{censor}
\usepackage{latexsym,fancyhdr,url}
\usepackage{enumerate}
\usepackage{algpseudocode}
\usepackage{graphics}
\usepackage{xparse} 
\usepackage{xspace}
\usepackage{multirow}
\usepackage{csvsimple}
\usepackage{balance}
\usepackage{pifont}
\usepackage{soul}
\usepackage{pgfplots}
\usepackage{listings}
\usepackage{xcolor}
\usepackage{xurl}
\usepackage[normalem]{ulem}
\usepackage[linesnumbered,ruled,vlined]{algorithm2e}

\definecolor{codegreen}{rgb}{0,0.6,0}
\definecolor{codegray}{rgb}{0.5,0.5,0.5}
\definecolor{codepurple}{rgb}{0.58,0,0.82}
\definecolor{backcolour}{rgb}{0.95,0.95,0.92}
\definecolor{darkgreen}{rgb}{0.0, 0.5, 0.0}
\definecolor{ggreen}{HTML}{097969}
\definecolor{ggreen1}{HTML}{9DEC9D}
\definecolor{ppurple}{HTML}{9F4C7C}
\definecolor{yyellow}{HTML}{FFC000}

\lstdefinestyle{mystyle}{
  backgroundcolor=\color{backcolour}, commentstyle=\color{codegreen},
  keywordstyle=\color{magenta},
  numberstyle=\tiny\color{codegray},
  stringstyle=\color{codepurple},
  basicstyle=\ttfamily\footnotesize,
  breakatwhitespace=false,         
  breaklines=true,                 
  captionpos=b,                    
  keepspaces=true,                 
  numbers=left,                    
  numbersep=5pt,                  
  showspaces=false,                
  showstringspaces=false,
  showtabs=false,                  
  tabsize=2
}

\usepackage{
  tikz,
  pgfplots,
  pgfplotstable
}

\usetikzlibrary{
  shapes.geometric,
  arrows,
  external,
  pgfplots.groupplots,
  matrix
}
\newcommand{\tool}{{NoT.js}\xspace}
\newcommand{\adgr}{{AdGraph}\xspace}
\newcommand{\webgr}{{WebGraph}\xspace}
\newcommand{\pggr}{{PageGraph}\xspace}

\pgfplotsset{compat=1.9}

\newcommand{\ie}{\emph{i.e.,}\xspace}
\newcommand{\eg}{\emph{e.g.,}\xspace}

\usepackage{mathtools,}

\DeclareMathAlphabet{\mathcal}{OMS}{cmsy}{m}{n}
\definecolor{last-year}{HTML}{1E476C}
\definecolor{this-year}{HTML}{9FC5E8}
\newcommand{\inarxivversion}[1]{#1}

\DeclareGraphicsExtensions{%
    .png,.PNG,%
    .pdf,.PDF,
    .jpg,.mps,.jpeg,.jbig2,.jb2,.JPG,.JPEG,.JBIG2,.JB2}

\usepackage{xparse}
\newcommand{\bnm}{\begin{newmath}}
\newcommand{\enm}{\end{newmath}}

\newcommand{\bea}{\begin{eqnarray*}}%
\newcommand{\eea}{\end{eqnarray*}}%

\newcommand{\bne}{\begin{newequation}}
\newcommand{\ene}{\end{newequation}}

\newcommand{\bal}{\begin{newalign}}
\newcommand{\eal}{\end{newalign}}

\newenvironment{newalign}{\begin{align}%
\setlength{\abovedisplayskip}{4pt}%
\setlength{\belowdisplayskip}{4pt}%
\setlength{\abovedisplayshortskip}{6pt}%
\setlength{\belowdisplayshortskip}{6pt} }{\end{align}}

\newenvironment{newmath}{\begin{displaymath}%
\setlength{\abovedisplayskip}{4pt}%
\setlength{\belowdisplayskip}{4pt}%
\setlength{\abovedisplayshortskip}{6pt}%
\setlength{\belowdisplayshortskip}{6pt} }{\end{displaymath}}

\newenvironment{newequation}{\begin{equation}%
\setlength{\abovedisplayskip}{4pt}%
\setlength{\belowdisplayskip}{4pt}%
\setlength{\abovedisplayshortskip}{6pt}%
\setlength{\belowdisplayshortskip}{6pt} }{\end{equation}}

\newcounter{ctr}

\newcounter{mytable}
\def\mytable{\begin{centering}\refstepcounter{mytable}}
\def\endmytable{\end{centering}}

\newcounter{myfig}
\def\myfig{\begin{centering}\refstepcounter{myfig}}
\def\endmyfig{\end{centering}}

\newlength{\saveparindent}
\newlength{\saveparskip}
\newcommand{\E}{{\rm I\kern-.3em E}}

\renewcommand{\eqref}[1]{\mbox{Equation~(\ref{#1})}}

\def \part {part}

\renewcommand{\paragraph}[1]{\vspace*{6pt}\noindent\textbf{#1}\;}

\def \blackslug{\hbox{\hskip 1pt \vrule width 4pt height 8pt
    depth 1.5pt \hskip 1pt}}
\def \qed{\quad\blackslug\lower 8.5pt\null\par}

\newcounter{mynote}[section]

\newcommand{\thenote}{\thesection.\arabic{mynote}}

\newcommand\ignore[1]{}

\newcounter{rcnote}[section]

\newcounter{mrnote}[section]

\newcounter{fknote}[section]

\newcounter{anote}[section]

\DeclareMathSymbol{\mlq}{\mathord}{operators}{``}
\DeclareMathSymbol{\mrq}{\mathord}{operators}{`'}

\newcommand{\rhf}[2]{R_{f, \gamma}}

\DeclareDocumentCommand{\edist}{o o}{
  \ensuremath{
    \IfNoValueTF{#1}{{d}}{{\sf d}(#1,#2)}
  }
}

\newcommand{\olrk}[1]{\ifx\nursymbol#1\else\!\!\mskip4.5mu plus 0.5mu\left(\mskip0.5mu plus0.5mu #1\mskip1.5mu plus0.5mu \right)\fi}

\NewDocumentCommand{\indseq}{ O{1} O{r} }{{#1}\ldots {#2}}

\providecommand{\etal}{\emph{et al.}\xspace}

\providecommand{\UsePackageFor}[2]{ \ifx#2\undefined\usepackage{#1}\fi }

\UsePackageFor{amssymb}{\boxtimes}
\usepackage[normalem]{ulem}		
\usepackage{bbding}				
\usepackage{endnotes}			
\usepackage{hyperref}			
\usepackage{hyperendnotes}	
\usepackage{pdfcolfoot}			

\ifdefined\OrigFootnote\relax\else
	\newenvironment{FootnoteContent}{}{}
	\let\OrigFootnote\footnote
	\let\OrigFootnoteText\footnotetext
	\renewcommand{\footnotetext}[1]{\OrigFootnoteText{\begin{FootnoteContent}#1\end{FootnoteContent}}}
	\renewcommand{\footnote    }[1]{\OrigFootnote    {\begin{FootnoteContent}#1\end{FootnoteContent}}}
\fi

\definecolor{PurplePlum}{rgb}{0.1,0,0.55} 
\definecolor{Brown}{rgb}{0.5,.25,0}
\definecolor{Orange}{rgb}{1,.3,0}
\definecolor{Gray}{rgb}{.7,.7,.7}
\definecolor{DarkGreen}{rgb}{.1,.41,.1}

\newif\ifBleck

\newcommand\Colour[1] {\color{#1}}

\newcommand\PrintToCLinks{	
  {\Colour{blue}\mbox{
    \hyperlink{w1619}{\sf$\rightarrow$~top}\quad
    \hyperlink{w1031}{\sf$\rightarrow$~toc}\quad
    \hyperlink{w1148}{\sf$\rightarrow$~lof}\quad
    \hyperlink{GreenRoom}{\sf$\rightarrow$~gr}\quad
    \hyperlink{EndNotes}{\sf$\rightarrow$~en}\quad
    \hyperlink{Sargasso}{\sf$\rightarrow$~sg}\quad
    \hyperlink{Index}{\sf$\rightarrow$~idx}
  }}
}
\makeatletter 
\newcommand\ToCLinks{
  \ifx\@onlypreamble\@notprerr		
    \hypertarget{w1619}{}			
  \else
    \AtBeginDocument{\hypertarget{w1619}{}}	
  \fi

  \ifBleck\else	
    \ifdefined\cofoot
      \cofoot{\PrintToCLinks}
      \cefoot{\PrintToCLinks}
    \else
      \def\@oddfoot{\PrintToCLinks}
      \def\@evenfoot{\PrintToCLinks}
    \fi
 \fi
}
\makeatother

\newif\ifEndNotes 

\newcommand\FnSym{{\scriptsize\PencilLeftDown\kern.1em}}		
\newcommand\EnSym {{$\bigtriangledown$}}

\def\MarkupsHowto{} 
\newcommand{\MarkupsHowtoAdd}[1]{\expandafter\def\expandafter\MarkupsHowto\expandafter{\MarkupsHowto{}#1}} 

\newif\ifMarkupsHowtoPrinted 
\newif\ifSuppress 

\newcommand\MakeMarkups[3][.]{

     \Suppressfalse
     \ifBleck\Suppresstrue\fi
     \ifx0#1\Suppresstrue\fi
     \ifx1#1\Suppressfalse\fi
     
     \expandafter\providecommand\csname#2x\endcsname {} 
     \ifSuppress\expandafter\renewcommand\csname#2x\endcsname{\relax}\else
                       \expandafter\renewcommand\csname#2x\endcsname{#3}\fi
                       
     \expandafter\providecommand\csname#2\endcsname {} 
     \ifSuppress\expandafter\renewcommand\csname#2\endcsname[1]{##1}\else
                       \expandafter\renewcommand\csname#2\endcsname[1]{{\csname#2x\endcsname##1}}\fi

     \expandafter\providecommand\csname#2d\endcsname {} 
     \ifSuppress\expandafter\renewcommand\csname#2d\endcsname[1]{\relax}\else
                       \expandafter\renewcommand\csname#2d\endcsname[1]{{\csname#2x\endcsname\sout{##1}}}\fi
                       
     \expandafter\providecommand\csname#2r\endcsname {} 
     \ifSuppress\expandafter\renewcommand\csname#2r\endcsname[2]{{##2}}\else
                       \expandafter\renewcommand\csname#2r\endcsname[2]{\csname#2d\endcsname{##1} \csname#2\endcsname{##2}}\fi

     \expandafter\providecommand\csname#2i\endcsname {} 
     \ifSuppress\expandafter\renewcommand\csname#2i\endcsname[1]{\relax}\else
                       \expandafter\renewcommand\csname#2i\endcsname[1]{\csname#2\endcsname{##1}}\fi

     \expandafter\providecommand\csname#2t\endcsname {} 
     \ifSuppress\expandafter\renewcommand\csname#2t\endcsname[1]{\relax}\else
                       \expandafter\renewcommand\csname#2t\endcsname[1]{{\csname#2x\endcsname{\mbox{$\langle\!\langle$}##1{\csname#2x\endcsname\mbox{$\rangle\!\rangle$}}}}}\fi 

     \expandafter\providecommand\csname#2b\endcsname {} 
     \ifSuppress\expandafter\renewcommand\csname#2b\endcsname[1][empty]{\relax}\else 
                       \expandafter\renewcommand\csname#2b\endcsname[1][\empty]{\ifx\empty##1\empty
                       	\label{#2-bookmark} 
                              \marginpar [\raggedleft\csname#2\endcsname{{\footnotesize\fbox{#2 working here}}~$\Longrightarrow$}]
                                                {\csname#2\endcsname{$\Longleftarrow$~{\footnotesize\fbox{#2 working here}}}}
                       \else 
                       	\marginpar [\raggedleft\csname#2\endcsname{\ifx\empty##1\empty\else\fbox{\tiny\parbox{8em}{\raggedright##1}}~\fi$\Longrightarrow$}]
                                                {\csname#2\endcsname{$\Longleftarrow$\ifx\empty##1\empty\else~{\tiny\fbox{\parbox{8em}{\raggedright##1}}}\fi}}\fi}\fi

     \expandafter\providecommand\csname#2TD\endcsname {} 
     \ifSuppress\expandafter\renewcommand\csname#2TD\endcsname{\relax}\else
                       \expandafter\renewcommand\csname#2TD\endcsname{\csname#2\endcsname{\fbox{#2 to do}}}\fi

     \expandafter\providecommand\csname#2Bar\endcsname {} 
     \ifSuppress\expandafter\renewcommand\csname#2Bar\endcsname{\relax}\else
                       \expandafter\renewcommand\csname#2Bar\endcsname{\csname#2\endcsname{\scriptsize\XSolidBrush}}\fi

     \expandafter\providecommand\csname#2f\endcsname {} 
     \ifSuppress\expandafter\renewcommand\csname#2f\endcsname[2][]{\relax}\else
      \expandafter\renewcommand\csname#2f\endcsname[2][\empty]{ 
        {\mbox{\csname#2x\endcsname\tiny$\boxtimes$}\marginpar{\hsize1cm\csname#2x\endcsname\fbox{\FnSym\footnotemark}}\relax 
        \footnotetext{\csname#2x\endcsname##2}}}\fi

     \expandafter\providecommand\csname#2e\endcsname {}
     \ifSuppress\expandafter\renewcommand\csname#2e\endcsname[1]{\relax}\else%
      \expandafter\renewcommand\csname#2e\endcsname[1]{%
       \global\EndNotestrue
       \mbox{\scriptsize\csname#2x\endcsname$\boxtimes$}\relax%
       \marginpar{\hsize1cm\csname#2x\endcsname\fbox{\EnSym\endnotemark%
                          \hypertarget{ENmark\thepage-\theendnote}{}~\hyperlink{ENtext\thepage-\theendnote}{{\Colour{blue}$\downarrow$}}}%
       }%
       {
        \def\zz{\noexpand#3}%
        \edef\z{~{[Endnote \theendnote\ %
        on p.\noexpand\hypertarget{ENtext\thepage-\theendnote}{}\thepage%
                    ~\noexpand\hyperlink{ENmark\thepage-\theendnote}{{\noexpand\Colour{blue}$\uparrow$}}]}%
        }%
        \expandafter\endnotetext\expandafter{\z\vspace{2ex}\\ ##1\newpage}%
       }
      }\fi

     \expandafter\providecommand\csname#2n\endcsname {}
     \ifSuppress\expandafter\renewcommand\csname#2n\endcsname[1]{\relax}\else%
      \expandafter\renewcommand\csname#2n\endcsname[1]{%
       \global\EndNotestrue
    \marginpar{{\tiny\endnotemark}\hypertarget{ENmark\thepage-\theendnote}{}~\hyperlink{ENtext\thepage-\theendnote}{}}
       {
        \def\zz{\noexpand#3}%
        \edef\z{~{\zz[Endnote (deferred) 
        from p.\noexpand\hypertarget{ENtext\thepage-\theendnote}{}\thepage%
        ]}%
        }%
        \expandafter\endnotetext\expandafter{\z\vspace{2ex}\\ ##1\newpage}%
       }
      }\fi

     \expandafter\providecommand\csname#2fe\endcsname {} 
     \ifSuppress\expandafter\renewcommand\csname#2fe\endcsname[2][]{\relax}\else 
      \expandafter\renewcommand\csname#2fe\endcsname[2][]{ 
       \def\File{##1}\relax
       \ifx\File\empty\csname#2f\endcsname{##2}\else 
        \global\EndNotestrue 
        \mbox{\scriptsize\csname#2x\endcsname$\boxtimes$}
        \marginpar{\csname#2x\endcsname\fbox{\FnSym\footnotemark}}\relax
        \footnotetext{~\csname#2x\endcsname##2\
                             --- See [\EnSym\endnotemark\hypertarget{ENmark\thepage-\theendnote}{}
                             \kern-.2em\hyperlink{ENtext\thepage-\theendnote}{{\Colour{blue}$\downarrow$}}].}\relax
       { 
         \def\zz{\noexpand#3}
         \edef\z{~{\zz[Endnote~\thefootnote~on~p.\noexpand\hypertarget{ENtext\thepage-\theendnote}{}\thepage
                     ~\noexpand\hyperlink{ENmark\thepage-\theendnote}
                     {{\noexpand\Colour{blue}\kern-0.1em$\uparrow$}]}}
                     {\noexpand\footnotesize\noexpand\newline\noexpand\hspace*{2em} (~from file {\noexpand\tt\File.tex}~)}
         }    
         \expandafter\endnotetext\expandafter{\z~\par\input{##1}\newpage}
        } 
       \fi 
      } 
     \fi 

     \ifSuppress\relax\else\ifBleck\relax\else
      \MarkupsHowtoAdd{\par\csname#2t\endcsname{
       $\backslash$\texttt{#2}$\cdots$\ markups are in \textbf{this} colour\ifx#1..\else\ifx1#1.\else, e.g.\ for #1.\fi\fi
       \ifMarkupsHowtoPrinted\relax\else 
        \global\MarkupsHowtoPrintedtrue 
        \begin{quote}\begin{tabular}{l@{\hspace{2em}}p{.7\linewidth}}
         \multicolumn{2}{l}{\texttt{$\backslash$MakeMarkups\ifx#1.\relax\else[#1]\fi\{#2\}\{{\it$\langle$colour command\/$\rangle$}\}}
         				 --- Defines the macros below:}\\
             & see comments at \texttt{$\backslash$MakeMarkups} definition. \\[1ex]
         \texttt{$\backslash$#2\{$\langle$text$\rangle$\}} & Sets \texttt{$\langle$text$\rangle$} in \texttt{#2}'s colour. \\
         \texttt{$\backslash$#2x} & Changes to \texttt{#2}'s colour (until end of context). \\
         \texttt{$\backslash$#2d\{$\langle$text$\rangle$\}} & Sets \texttt{$\langle$text$\rangle$} in \texttt{#2}'s colour with a strikethrough (i.e.\ delete). \\
         \texttt{$\backslash$#2r\{$\langle$this$\rangle$\}\{$\langle$that$\rangle$\}} &
          Strikes through \texttt{$\langle$this$\rangle$} and inserts \texttt{$\langle$that$\rangle$} (i.e.\ replace). \\
         \texttt{$\backslash$#2f\{$\langle$text$\rangle$\}} & Meta-comment: puts \texttt{$\langle$text$\rangle$} in a \texttt{#2}-footnote with a {\tiny$\boxtimes$} in the main text. \\
         \texttt{$\backslash$#2t\{$\langle$text$\rangle$\}} & Use for meta when  \texttt{$\backslash$#2f} isn't allowed (``Not in outer-par mode.'') \\
         \texttt{$\backslash$#2b[$\langle$optional$\rangle$]} & Marginal pointer, with label for hyper-linking directly there. \\
         \texttt{$\backslash$#2e\{$\langle$text$\rangle$\}} & Puts \texttt{$\langle$text$\rangle$} in a \texttt{#2}-endnote with a (big) $\boxtimes$ in the main text. \\[.5ex]
         \texttt{$\backslash$#2n\{$\langle$text$\rangle$\}} & Like \texttt{$\backslash$#2e}
         except there's no reference from the main text. Good for ``decluttering''
         when you still want to have the footnote- or endnote texts as reminders. \\[.5ex]
         \texttt{$\backslash$#2fe[$\langle$this$\rangle$]\{$\langle$that$\rangle$\}} & Makes a \texttt{$\backslash$#2f\{$\langle$that$\rangle$\}} that refers to a \\
           & \texttt{$\backslash$#2e\{$\langle$contents of file this.tex$\rangle$\}}. \\ 
           & Without the optional argument, acts as \texttt{$\backslash$#2f\{$\langle$that$\rangle$\}}. \\[.5ex]
         \texttt{$\backslash$#2Bar} & Inserts ``burn after reading'' symbol \csname#2Bar\endcsname, meaning
          \begin{quote}\begin{itemize}\setlength\itemsep{0pt}
           \item If yours is the only \csname#2Bar\endcsname\ in this (presumably someone else's) footnote, and you are happy that the footnote has been addressed,
           go ahead and comment-out the whole footnote. (The \csname#2Bar\endcsname\ is their request for you to ``approve and remove''.)
           \item If you are not happy, delete only your \csname#2Bar\endcsname\ and follow-on in the footnote
            (in your colour, i.e.\ with \texttt{$\backslash$#2x}) saying why you are not happy.
           \item If you are happy, but there are others' burn-after-reading symbols as well as yours, just delete yours; the other people have not yet responded.
          \end{itemize}
          \end{quote}
          The idea is that when everyone's happy, the last person will comment-out the meta-text. \\[0.5ex]
         \texttt{$\backslash$#2TD} & Inserts {\csname#2TD\endcsname}\ . \\
        \end{tabular}\end{quote}
       \fi
      }}
     \fi\fi
}

\newif\ifNoGreenRoom
\newcommand\MakeGreenRoom {\ifBleck\relax\else\ifNoGreenRoom\relax\else
\newcommand\NewGRLabel[1] {\OldGRLabel{GreenRoom-##1}} 
 \newcommand\NewGRRef[1] 
 {\expandafter\ifx\csname r@GreenRoom-##1\endcsname\relax\OldGRRef{##1}\else\OldGRRef{GreenRoom-##1}\fi}
 \let\OldGRLabel\label \let\label\NewGRLabel
 \let\OldGRRef\ref \let\ref\NewGRRef
 \hrule
 ~\\\begin{center}\Huge \hypertarget{GreenRoom}{Green Room}
 \end{center}~\\
 \hrule
\fi\fi}

\newcommand\EndGreenRoom  {\ifBleck\relax\else\ifNoGreenRoom\relax\else
\let\label\OldGRLabel
\let\ref\OldGRRef
\fi\fi}

\newif\ifNoEndNotes

\newif\ifNoSargasso
\newcommand\MakeSargasso {
 \hypertarget{Sargasso}{}
 \newcommand\NewLabel[1] {\OldLabel{Sargasso-##1}} 
 \newcommand\NewRef[1] 
 {\expandafter\ifx\csname r@Sargasso-##1\endcsname\relax\OldRef{##1}\else\OldRef{Sargasso-##1}\fi}
 \let\OldLabel\label \let\label\NewLabel
 \let\OldRef\ref \let\ref\NewRef
\ifBleck\end{document}\else\ifNoSargasso
\relax
\else
  \hrule
  ~\\\begin{center}\Huge Sargasso
  \end{center}~\\
  \hrule
 \fi\fi
}

\newcommand\EndSargasso  {\ifBleck\relax\else\ifNoSargasso\relax\else
\let\label\OldLabel
\let\ref\OldRef
\fi\fi}

\newcommand\EndDocument {\ifBleck\end{document}\fi} 

\newcommand\Cite[2][\empty] {{\Colour{red}\ifx#1\empty[#2]\else[#2,~#1]\fi}}

\begin{document}

\def\thetitle{Blocking Tracking JavaScript at the Function Granularity}
\def\thenote{Thanks to the XYZ library for their support.}



\title{\thetitle}

\author{
{\rm Abdul Haddi Amjad \textsuperscript{*} \ \ \
     Shaoor Munir\textsuperscript{\textdagger} \ \ \
     Zubair Shafiq\textsuperscript{\textdagger} \ \ \
     Muhammad Ali Gulzar\textsuperscript{*}
}
\ \ \ \ \ \ \ \ \ \ \ \ \ \ \
{\rm \textsuperscript{*}Virginia Tech, Blacksburg} \ \ \
{\rm \textsuperscript{\textdagger}University of California, Davis}
\ \ \ \ \ \ \ \ \ \ \ \ \ \ \ \ \ \ \ \ \ \ \ \ \ \ \ \ \ \ \ \ \ \ \ \ \ \ \ \ \ \ \ \ \ \ \ \ \ \ \ \ \ \ \ \ \ \ \ \
\inarxivversion{
{\rm \textcolor{blue}{This is an extended version of our paper that appears in ACM CCS 2024}}
}
} %

\renewcommand{\shortauthors}{Amjad et al.}
\date{}
\newcommand\name{MethodFlow\xspace}
\newcommand\webgraph{WebGraph\xspace}

\MakeMarkups[Hadi]{U}{\Colour{blue}}
\MakeMarkups[Shaoor]{S}{\Colour{DarkGreen}}
\MakeMarkups[Zubair]{Z}{\Colour{magenta}}
\MakeMarkups[Gulzar]{G}{\Colour{orange}}

\begin{abstract}
Modern websites extensively rely on JavaScript to implement both functionality and tracking. 
Existing privacy-enhancing content blocking tools struggle against mixed scripts, which simultaneously implement both functionality and tracking, because blocking the script would break functionality and not blocking it would allow tracking.
We propose \tool, a fine-grained JavaScript blocking tool that operates at the function-level granularity. 
\tool's strengths lie in analyzing the dynamic execution context, including the call stack and calling context of each JavaScript function, and then encoding this context to build a rich graph representation.
\tool trains a supervised machine learning classifier on a webpage's graph representation to first detect tracking at the JavaScript function-level and then automatically generate surrogate scripts that preserve functionality while removing tracking. 
Our evaluation of \tool on the top-10K websites demonstrates that it achieves high precision (94\%) and recall (98\%) in detecting tracking JavaScript functions, outperforming the state-of-the-art while being robust against off-the-shelf JavaScript obfuscation. 
Fine-grained detection of tracking functions allows \tool to automatically generate surrogate scripts that remove tracking JavaScript functions without causing major breakage.
Our deployment of \tool shows that mixed scripts are present on 62.3\% of the top-10K websites, with 70.6\% of the mixed scripts being third-party that engage in tracking activities such as cookie ghostwriting. 
We share a sample of the tracking functions detected by \tool within mixed scripts—not currently on filter lists—with filter list authors, who confirm that these scripts are not blocked due to potential functionality breakage, despite being known to implement tracking.

%
\end{abstract}


\maketitle

\section{Introduction}
\label{sec:intro}
Modern websites extensively rely on third-party JavaScript (JS) to implement tracking (\eg advertising, analytics) and non-tracking (\eg functional content) \cite{JavaScript2022WebAlmanac, chen2021cookie, mayer2012third}. 
In fact, 81\% of all tracking requests are triggered by JS \cite{amjad2023blocking}. 
Privacy-enhancing content blockers aim to block tracking JS without breaking the legitimate website functionality. 
As the arms race has escalated with trackers attempting to evade blocking, the state-of-the-art content blocking approaches face the following key challenges in blocking tracking JS \cite{ubo-web,Iqbal20AdgraphSP,Siby22WebGraph,FilterlistGeneration2020ACMWWW,Chen21EventLoopS&P}.

First, the state-of-the-art content blocking approaches do not capture the context needed to effectively detect tracking behavior implemented in a script.
While existing approaches (\eg \cite{Siby22WebGraph,Chen21EventLoopS&P,Amjad2021TrackerSift}) capture the script that initiates a network request, they do not capture 
the sequence of functions that led to the network request. 
Such code-level interactions leading to network requests can provide rich clues to a web page's intent. 
The lack of sufficient consideration of the execution context associated with tracking behavior fundamentally limits the effectiveness of existing approaches. 

Second, the state-of-the-art content blocking approaches struggle when tracking and non-tracking resources are lumped together \cite{senol2024double,Amjad2021TrackerSift}.
When both tracking and non-tracking resources are served from the same network location, filter lists \cite{easylist,easyprivacy} or even ML approaches \cite{Iqbal20AdgraphSP,Siby22WebGraph} struggle.
This issue is further exacerbated by the use of URL encryption \cite{facebookurlobfuscation,wang2016webranz}.
This means that URL-based approaches are unable to discern between tracking and non-tracking resources.
Similarly, JS signatures \cite{Chen21EventLoopS&P} are ineffective when tracking and non-tracking code is combined in the same script.  
This is fundamentally a granularity issue -- \ie specific functions within mixed scripts are responsible for tracking \cite{Amjad2021TrackerSift}.

Third, the state-of-the-art content blocking approaches rely on the laborious process of manually curated filter lists.
Filter lists are used to block tracking network requests \cite{easylist,easyprivacy} and stop the execution of tracking code \cite{adguardsnippets, repl}. 
TrackerSift \cite{Amjad2021TrackerSift} relies on manually curated filter lists to detect mixed tracking scripts.  
uBlock Origin employs manually refactored replacements (aka scriptlets) to handle mixed tracking scripts \cite{ubo-scriptlets,GoogleAnalyticsReplacement}. 
The reliance on manual curation fundamentally hinders scalability.

\tool advances the state-of-the-art by addressing these three challenges. 
First, \tool leverages browser instrumentation to capture dynamic execution context, including the call stack and the calling context of each function call in the call stack. While a JS function's static representation remains unchanged, the execution context around it may alter its semantics. Dynamic execution context enables \tool to semantically reason about a JS function execution, which is essential in differentiating its participation in tracking and non-tracking activity. 
Second, \tool leverages this dynamic execution context to encode fine-grained JS execution behavior in a rich graph representation that includes individual JS functions within each script. 
Third, \tool trains a supervised machine learning classifier to detect tracking at the function-level granularity and automatically generate surrogate scripts to specifically block the execution of tracking functions while not impacting execution of non-tracking functions.
\tool is the first to fully automate the entire surrogate generation process end-to-end, which are currently painstakingly hand-crafted by experts.

We evaluate the effectiveness, robustness, and usability of \tool on the top-10K websites from the Tranco list \cite{LePochat2019}.
Our evaluation shows that \tool accurately detects tracking JS functions with 94\% precision and 98\% recall, outperforming the state-of-the-art by up to 40\%. \tool's contributions in incorporating dynamic execution context account for 29\% improvement in F1-score. 
Against a number of JS obfuscation techniques, such as control flow flattening, dead code injection, functionality map, and bundling, \tool remains fairly robust -- its F1-score decreases by only 4\%.
\tool's automatically generated surrogate scripts block 84\% of the tracking JS function calls without causing any breakage on 92\% of the websites.
%

We deploy \tool to study the tracking functions in mixed scripts, discovering that 62\% of the top-10K websites have at least one mixed script. 
We find that the tracking functions are served in the mixed scripts from more than eight thousand unique domains, including those belonging to tag management services, advertisers, and content delivery networks (CDNs). 
Notably, among these mixed scripts, a significant 70.6\% are third-party scripts that engage in tracking activities such as cookie ghostwriting \cite{Sanchez2021JourneyToCookies,munir2023cookiegraph}.

Our key contributions are summarized as follows:
\begin{enumerate}

\item We propose \tool, a machine learning-based approach to detect and block tracking at the JS function-level granularity. We show that \tool outperforms the state-of-the-art in terms of accuracy and is robust against evasion. 

\item We implement \tool as a browser extension and show that it can be used to automatically generate surrogate scripts by neutralizing tracking function calls. We show that these surrogate scripts can be injected into a website to reliably mitigate tracking at its origin without breaking website functionality. 

\item We report a sample\footnote{EasyList's \href{https://github.com/easylist/easylist/issues/new}{"issue-protocol"} explicitly warns against frequent creation or reopening of issues to avoid being banned.} of mixed scripts that are detected by \tool (as containing both tracking and non-tracking functions) but missed by the filter list authors \cite{easylist,easyprivacy}. 
The filter list authors confirm (via Github issues \cite{18230,18242,18185, 18243}) that these mixed scripts are not blocked by filter lists despite being known to implement tracking due to breakage concerns.

    
\item We deploy \tool on the top-10K websites to measure the prevalence of tracking functions in the mixed scripts. We show that these mixed scripts are commonly served by tag management services, advertisers, and content delivery networks (CDNs). A majority of these mixed scripts are third-party, actively engaged in tracking activities such as cookie ghostwriting.

\end{enumerate}

\vspace{.05in} \noindent {\bf Data Availability:} Our code and data are available at
\href{https://zenodo.org/search?q=parent.id%3A10287302&f=allversions%3Atrue&l=list&p=1&s=10&sort=version}{\nolinkurl{https://zenodo.org/search?q=NoT.JS}}.    
\section{Background \& Related Work}
\label{sec:relwork}

\noindent\textbf{Existing countermeasures against web tracking.}
Privacy enhancing content blockers, such as uBlock Origin \cite{ubo-web} and Brave \cite{brave-tracking-protection}, primarily rely on manually curated filter lists \cite{easylist,easyprivacy,edge-tracking-protection-doc,brave-tracking-protection,firefox-etp} to block tracking network requests at the client side.
These filter lists contain URL- or domain-based rules to determine whether a network request should be allowed or blocked.
Prior research shows that trackers are able to evade filter lists by changing their network location, such as the URL or domain \cite{DGA2018BleepingComputer, lin2022investigating, mughees2016first, vadrevu2019you}.
These evasions necessitate manual updates to the filter lists to accommodate the new network locations, leading to a perpetual arms race between the maintainers of filter lists and trackers \cite{stealing2018netlab, adblockerassualtfacebook2017adage, controlfacebook2016newsroom, pingpongfacebook2016adblockplus, FilterFilters2020ACM, AdWars2017ACMIMC, WebRanz2016ACMSIGSOFT}.

To mitigate this issue, recent approaches, such as \adgr \cite{Iqbal20AdgraphSP}, \webgr \cite{Siby22WebGraph}, and \pggr \cite{FilterlistGeneration2020ACMWWW}, use machine learning to automatically generate filter list rules, aiming at the identification of tracking network requests.
These approaches adopt a graph-based representation of the webpage execution to classify network requests.
Both \webgr and \pggr enhance \adgr by incorporating additional features into their graph representation.
For instance, \webgr additionally captures storage accesses (\eg cookie read/write), exhibiting superior performance against URL/domain-based evasion.
While these approaches detect and block tracking network requests, 
%
their instrumentation is limited to the interactions between the initiator script and the tracking request, making it ineffective in pinpointing the origin of tracking activity within mixed resources.

\noindent\textbf{Script-level restrictions against web tracking.}
Trackers/advertisers evade content blockers by utilizing a common network location to serve both tracking and non-tracking resources.
For instance, trackers have started using Content Delivery Networks (CDNs) or engage in CNAME cloaking \cite{Dimova21CNAMEPETS, Dao2021CNAMECloaking, dimova2021cname} to serve both tracking and non-tracking requests from a common network location.
Content blockers are thus presented with a dilemma: either block the network request, potentially disrupting legitimate website functionality, or permit the network request, consequently letting go of tracking/advertising.
To address this issue, a few content blockers such as uBlock Origin have introduced support for ``scriptlets", which are custom JS snippets injected at runtime to substitute the code that initiates tracking/advertising network requests \cite{ubo-scriptlets}.
This scriptlet strategy is effective in countering tracking, even when originating from the same network location as non-tracking resources, as it eliminates tracking at its origin --- well before a tracking network request is initiated.
However, akin to filter lists, scriptlets are manually crafted and, therefore, challenging to scale across the entire web. 
At present, Brave Browser and uBlock Origin are capable of blocking a mere 27 scripts\footnote{\url{https://github.com/gorhill/uBlock/tree/master/src/web_accessible_resources}}, while popular filter lists comprise over 6,000 exception rules designed to enable functionality-critical tracking scripts \cite{snyder2020filters}. 
Consequently, tens of thousands of privacy-invasive scripts remain unblocked.

\begin{figure}[t]
    \centering
      \includegraphics[width=0.30\textwidth]{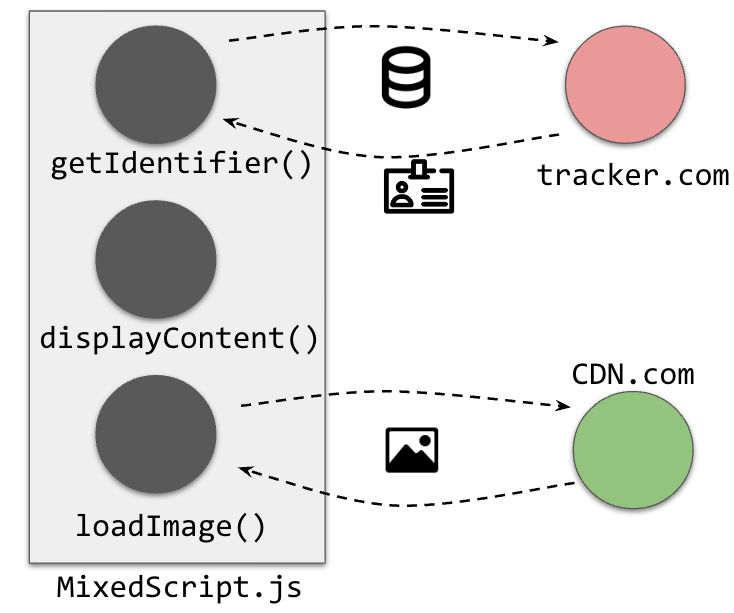}
      \put(-83,111){{\ding{202}}}
      \put(-83,80){{\ding{203}}}
      \put(-83,35){{\ding{204}}}
      \put(-83,13){{\ding{205}}}
      \vspace{-0.05in}
        \caption{Marker \ding{202} signifies a tracking request to {\tt tracker.com}, returning a user ID (\ding{203}) for activity monitoring, while \ding{204} indicates a non-tracking CDN request retrieving an image (\ding{205}).}
        \label{fig:mixed-script}
        \vspace{-0.10in}
\end{figure}
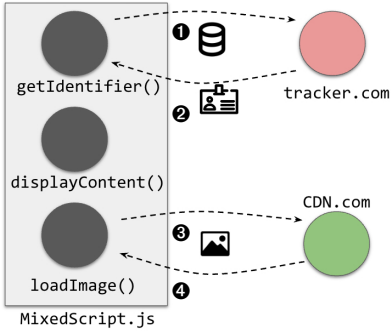

Static or dynamic program analysis approaches have also been proposed to detect tracking/advertising JS code. 
Ikram \etal \cite{ikram2016seamless} employed machine learning to analyze syntactical and structural aspects of JS code, aiming to classify tracking scripts.
However, their static analysis approach remains vulnerable to basic JS obfuscation techniques.
Chen \etal \cite{Chen21EventLoopS&P} devised event-loop-based signatures based on dynamic code execution to detect tracking scripts.
%
They found that some trackers bundle tracking and non-tracking code within a single script, posing a similar challenge when tracking and non-tracking resources are served from the same network location. 
The increasing prevalence of mixed scripts, particularly those facilitated through bundling, poses a fundamental challenge to privacy-enhancing content blocking \cite{Webbundles2020Brave}.
According to the Web Almanac, bundling is already a common practice on top-ranked websites; specifically, 17\% of the top-1K websites employ the Webpack JS bundler \cite{JavaScript2022WebAlmanac}. 
Furthermore, there has been a 5$\times$ increase in the downloads of prominent bundlers such as Webpack over the past five years \cite{NPM_trends}.
Amjad \etal showed that the prevalence of mixed scripts on top-100K websites increased from 12.8\% in 2021 \cite{Amjad2021TrackerSift} to 14.6\% in 2022 \cite{amjad2023blocking}.

\noindent\textbf{Function-level restrictions against web tracking.} 
Prior research has proposed a function-level characterization of JS code.
Smith \etal proposed SugarCoat \cite{smith2021sugarcoat} to systematically generate substitutes for scripts involved in tracking activities.
SugarCoat relies on existing filter lists to identify tracking scripts for rewriting. It is ineffective against false negatives present in these filter lists.
It uses \pggr \cite{FilterlistGeneration2020ACMWWW} to pinpoint the code locations where scripts access the privacy-sensitive data, such as {\tt document.cookie} and {\tt localStorage}.
Subsequently, these code locations, which can include JS functions, are replaced by benign mock implementations that are manually generated by developers.
However, to date, only six mock implementations \footnote{\url{https://github.com/SugarCoatJS/sugarcoat/tree/master/mocks}} are created for the designated APIs through developer assistance.
This reliance on manual effort from developers limits SugarCoat's applicability to a large number of tracking scripts.

Amjad \etal \cite{Amjad2021TrackerSift, amjad2023blocking} proposed TrackerSift, an approach to untangle mixed JS code down to the granularity of individual functions.
%
Their approach is able to separate out 98\% of all requests generated by tracking and non-tracking resources.
%
Their empirical analysis determines that targeting specific JS functions reduces breakage by 3.8$\times$ as compared to blocking entire scripts.
Nevertheless, TrackerSift's efficacy solely relies on existing filter lists for the differentiation of tracking and non-tracking resources, thus limiting its usefulness.


\vspace{-0.1in}
\section{Threat Model}
\label{sec:threatmodel}
In this section, we describe the threat model for mixed scripts -- JS that combines both tracking and non-tracking functionality, making it challenging for privacy-enhancing content blockers to detect and block them.

\begin{figure}[t]
    \centering
      \includegraphics[width=0.40\textwidth]{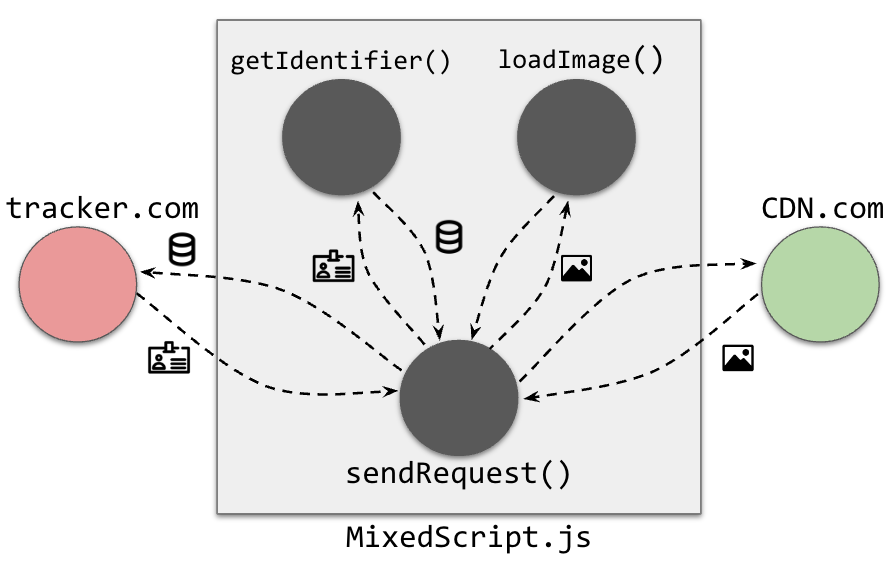}
      \put(-114,80){{\ding{202}}}
      \put(-118,67){{\ding{205}}}
      \put(-95,72){{\ding{206}}}
      \put(-85,68){{\ding{209}}}
      \put(-145, 65){{\ding{203}}}
      \put(-145,43){{\ding{204}}}
      \put(-60,67){{\ding{207}}}
      \put(-60,45){{\ding{208}}}
        \caption{ Marker \ding{202} signifies a call from {\tt getIdentifier} to {\tt sendRequest}, starting a tracking request to tracker \ding{203}, and the identifier received at \ding{204} is returned to the originating function at \ding{205}. Concurrently, \ding{206} represents a call from {\tt loadImage} to {\tt sendRequest}, triggering a non-tracking request \ding{207} to a CDN, with the image displayed at \ding{208} and returned to {\tt loadImage} at \ding{209}.}
        \label{fig:mixed-script-initiator}
        \vspace{-0.15in}
\end{figure}

\noindent\textbf{Definitions.}
%
We use the term {\tt initiator function} to describe a JS function that directly initiates a network request. 
This function is always at the top of the call stack when we analyze a network request.
In Figure \ref{fig:mixed-script}, both {\tt getIdentifier} and {\tt loadImage} are examples of initiator functions. 
Next, we use the term {\tt gateway function} to describe a specific type of initiator function that only initiates network requests and performs no other task. 
A gateway function essentially initiates network requests on behalf of other functions.
In Figure \ref{fig:mixed-script-initiator}, {\tt sendRequest} is an example of a gateway function.
Finally, we use the term {\tt "neutralize"} to remove tracking in a JS by replacing a tracking function call with a mock function call.

Below, we describe two specific techniques that trackers use to combine tracking and non-tracking JS code.

\noindent\textbf{Distinct tracking and non-tracking JS functions in the same script.} 
Figure \ref{fig:mixed-script} illustrates a script containing both tracking and non-tracking JS functions, each with separate roles.
In this example, {\tt getIdentifier} gathers user data and sends a tracking request to {\tt tracker.com} \ding{202}, which returns a user identifier \ding{203}.
This identifier is used to track user activity on the webpage.
The same script also includes functions like {\tt displayContent} and {\tt loadImage}, essential for the webpage to function properly. 
For instance, {\tt loadImage} sends a non-tracking request \ding{204} to {\tt CDN.com} to load an image \ding{205}.
Blocking the script to stop {\tt getIdentifier} would also disable essential functions like {\tt loadImage}, harming the webpage's functionality. 
Therefore, an ideal approach would neutralize tracking function calls to {\tt getIdentifier} while leaving the rest of the script untouched.

\begin{figure*}[t]
    \centering
      \includegraphics[width=\textwidth]{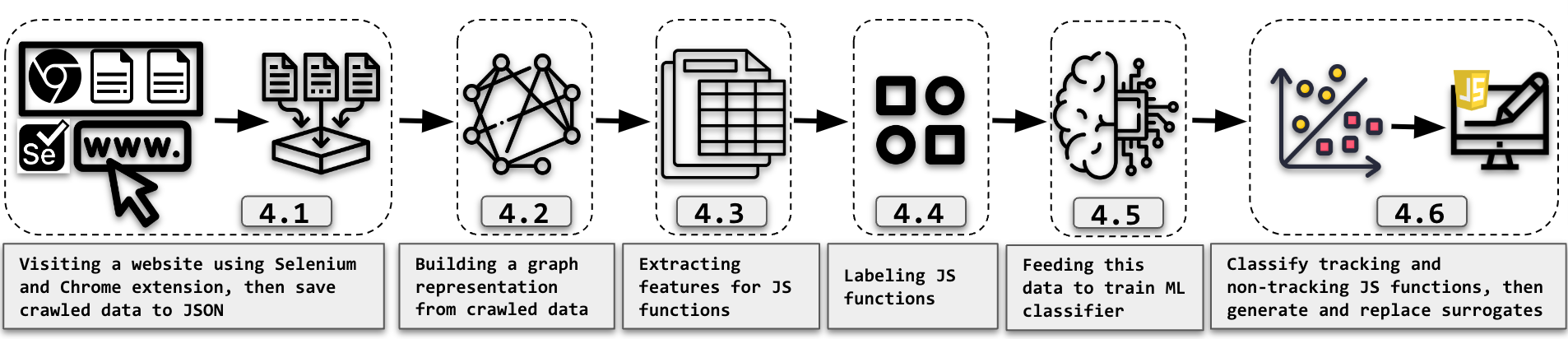}
       \put(-392,5){{\ding{202}}}
       \put(-316,5){{\ding{203}}}
      \put(-250,5){{\ding{204}}}
      \put(-193,5){{\ding{205}}}
      \put(-126,5){{\ding{206}}}
      \put(-9,5){{\ding{207}}}
        \caption{\tool pipeline: \ding{202} Crawl websites, save data; \ding{203} Create JS function graph; \ding{204} Extract features and \ding{205} label it; \ding{206} Train classifier; \ding{207} Classify tracking/non-tracking JS functions, and create surrogates.}
        \label{fig:overview}
\end{figure*}
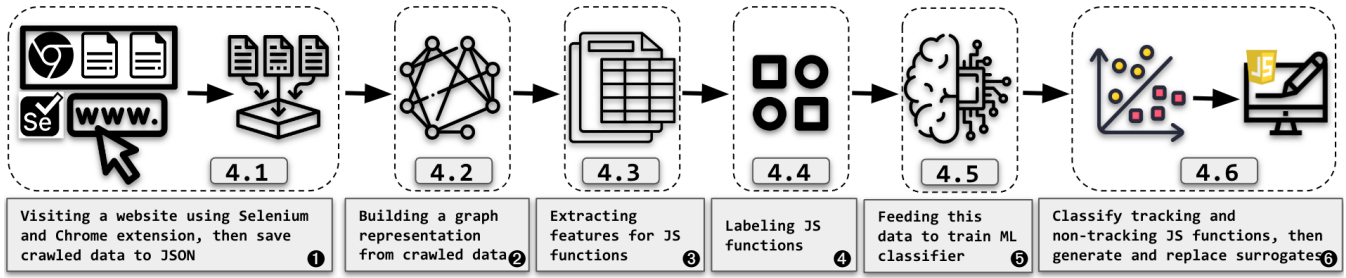
\noindent\textbf{Use of gateway functions to initiate tracking and non-tracking requests.}
Figure \ref{fig:mixed-script-initiator} illustrates how a gateway function, {\tt sendRequest}, is used to handle both tracking and non-tracking network requests.
The single script includes functions for both tracking and non-tracking tasks. 
Specifically, {\tt getIdentifier} gathers user data for {\tt tracker.com} to obtain a user identifier.
Conversely, {\tt loadImage} fetches images for the webpage from {\tt CDN.com}.
In contrast with the previous scenario, these functions do not initiate requests directly and delegate this task to the gateway {\tt sendRequest} function.
For instance, {\tt getIdentifier} calls {\tt sendRequest} \ding{202} to send request to {\tt tracker.com} \ding{203}.
In return, {\tt tracker.com} provides a user ID \ding{204}, which is sent back to {\tt getIdentifier} \ding{205}.
Similarly, {\tt loadImage} calls {\tt sendRequest} to fetch images from {\tt CDN.com}, which are then displayed on the webpage in \ding{206} - \ding{208}.

The gateway functions further abstract tracking and non-tracking in JS code and necessitate careful analysis of the execution context of a JS function to determine the most effective strategy to block the execution of tracking JS code.
%
Since the tracking function {\tt getIdentifier} is not always at the top of the call stack (Figure \ref{fig:mixed-script-initiator}), simply neutralizing the initiator function is not effective. 
Therefore, an ideal approach needs to incorporate a complete execution call stack and calling context to identify tracking JS functions.
In this case, only the function calls to {\tt sendRequest} invoked from {\tt getIdentifier} should be neutralized, which preserves the functionality when {\tt sendRequest} is invoked from {\tt loadImage}.

\section{\tool Design and Implementation}
\label{sec:methodology}

In this section, we describe the design and implementation details of \tool. 
Figure \ref{fig:overview} provides an overview of \tool's pipeline, which starts with automated website crawling using the Selenium web driver and collecting data using a Chrome browser instrumented with a custom-built extension \ding{202}.
Using the collected data, \tool generates a graph representation \ding{203} of each webpage that encodes JS dynamic execution context of a comprehensive list of webpage's activities like network requests, DOM modifications, storage access, and a subset of other Web APIs (listed in Table \ref{table:webapis}), that are commonly used by trackers \cite{bahrami2022fp}.
Using this graph representation, \tool extracts unique structural and contextual features of tracking activity \ding{204} and labels it \ding{205}, which are then utilized to train a random forest classifier capable of accurately identifying tracking entities \ding{206}.
Finally, the classification results are utilized to generate surrogate scripts by neutralizing tracking function calls. These surrogate scripts can be used as replacements by content blockers at runtime \ding{207}.

\subsection{\tool's Chrome Instrumentation}

\begin{figure}[t]
    \centering
      \includegraphics[width=0.40\textwidth]{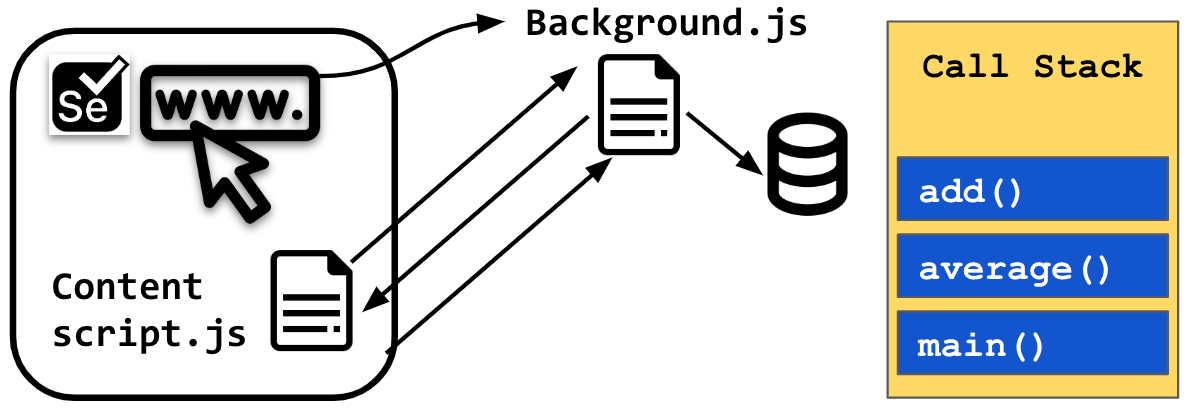}
       \put(-130,44){{\ding{202}}}
       \put(-125,58){{\ding{203}}}
       \put(-130,32){{\ding{204}}}
       \put(-130,22){{\ding{205}}}
       \put(-78,50){{\ding{206}}}
       \vspace{-0.05in}
        \caption{An illustration of \tool's chrome-extension, showing the communication sequence between background and content scripts.}
        \label{fig:extension}
        \vspace{-0.1in}
\end{figure}
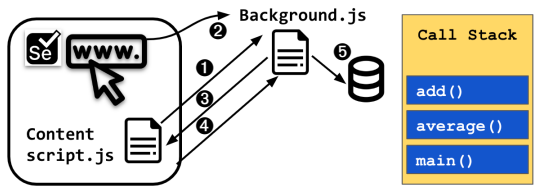

\tool first collects the training data to train a fine-grained, high-accuracy classifier. It automates the website crawling and data collection process using selenium \cite{selenium} and a custom-built Chrome extension. 
We choose Chrome extension interface \cite{chrome-extension} to capture web activities due to its ease of use and lightweight nature compared to instrumenting the JS engine, such as V8 in Chrome \cite{V8}.
\tool's Chrome extension captures the JS dynamic execution context for each activity on the webpage, along with other relevant meta-data such as network requests and response payloads. \tool's dynamic execution context for each web activity comprises of:  
\begin{itemize}
    \item The \textbf{call stack} \footnote{\url{https://chromedevtools.github.io/devtools-protocol/tot/Runtime/\#type-CallFrame}} outlines the sequence of function calls, including details like the script URL, method name, and line and column numbers where each function is invoked.
    \item The \textbf{scope chain} \footnote{\url{https://chromedevtools.github.io/devtools-protocol/tot/Debugger/\#type-Scope}} for each function call within the stack that includes the number of arguments and local and global variables.
\end{itemize}
The intuition behind capturing dynamic execution context is to gain a deeper understanding of the web activity. 
Figure \ref{fig:extension} illustrates the two primary components of the Chrome extension, namely the background script and content script, which work together to facilitate data collection.

\begin{table}[t]
\small
    \centering
    \scalebox{1}{\begin{tabular}{ |l|l|}
\hline
\textbf{Activity} & \textbf{Property}\\
\hline
    \multirow{4}{*}{Network Request}&Network.requestWillBeSent  \\
    & Network.responseReceived  \\
    & Network.requestWillBeSentExtraInfo  \\
    & Network.responseReceivedExtraInfo  \\
    \cline{1-2} 
    \multirow{3}{*}{DOM Modifications}& DOM.attributeModified  \\
    & DOM.childNodeInserted  \\
    & DOM.childNodeRemoved  \\
    \cline{1-2}
    \multirow{3}{*}{Storage Access}& Document.cookie (get/set)  \\
    & Storage.setItem  \\
    & Storage.getItem  \\
    \cline{1-2}
    \multirow{11}{*}{Web APIs}& Navigator.sendBeacon  \\
    & Navigator.geolocation  \\
    & Navigator.userAgent   \\     
    \cline{2-2}   
    & BatteryManager.chargingTime \\
    & BatteryManager.dischargingTime \\
    \cline{2-2}
    & MouseEvent.movementX  \\
    & MouseEvent.movementY   \\
    \cline{2-2}
    &Element.copy   \\
    &Element.paste    \\
    \cline{2-2}
    & Document.visibilitychange   \\
    \cline{2-2}
    &Touch.force   \\

\hline
\end{tabular}}
    \vspace{0.2in}
    \caption{List of webpage's activities captured by \tool.}
    \label{table:webapis}
    \vspace{-0.25in}
\end{table}

\noindent \textbf{Background script.}
The background script \cite{backgroundscript} is an essential component of the Chrome extension that captures webpage activity using the Chrome DevTools Protocol APIs \cite{chromedevtoolsChromeDevTools}. 
Specifically, the network API \cite{NetworkDomain} monitors traffic and provides valuable information about HTTP requests and responses, including headers, bodies, call stack, scope chain, and timestamps.
Additionally, the DOM API \cite{DOMDomain} captures changes in the Document Object Model and provides read and write operations along with the call stack, and the scope chain. 
However, the DevToolProtocol does not expose all Web APIs, cookies, or storage APIs, which are available through the content script as discussed next.

\noindent \textbf{Content script.}
The content script \cite{contentscript} runs within the context of the webpage and can interact with its functionality. 
It is responsible for collecting the JS dynamic execution context, \ie call stack and scope chain for the webpage's activity, and exposing Web APIs, cookies, and storage APIs that are not available in the background script, by overriding functions.
Listing \ref{lst:contentscript} shows the snippet from {\tt content.js} that overrides the {\tt sendBeacon} function of the {\tt Navigator} object.
The overridden function collects two types of information: the call stack of the JS execution and the scope chain of each function call in the call stack. 
The call stack is collected using the {\tt console.trace()} \cite{ConsoleTrace} function, which logs the stack trace. 
The scope chain is collected using the Debugger API \cite{DebuggerAPI}, specifically via the {\tt Debugger.paused} event that provides the {\tt callFrames.scopeChain} parameter.
\begin{lstlisting}[language=Java,label={lst:contentscript}, caption=Overriding {\tt sendBeacon} function using content script.]
// storing the original sendBeacon function
const sendBeac = Navigator.prototype.sendBeacon;
// overriding sendBeacon function 
Navigator.prototype.sendBeacon = function(url, data) {
    // collect stack trace and scope chain
    console.trace();
    Debugger.paused.callFrames.scopeChain;
    // call back the original function 
    sendBeac();}
\end{lstlisting}

Figure \ref{fig:extension} illustrates the sequence of the data collection process with \tool's chrome extension. 
First, {\tt content script.js} sets up communication with {\tt background.js}, as shown in step \ding{202}. 
On a network activity or DOM modifications (\ding{203}), {\tt background.js} triggers a message for {\tt content.js} to capture the corresponding JS execution call stack and scope chain, as shown in \ding{204}. 
Finally, the collected data is sent to {\tt background.js} for storage, as shown in \ding{205} and \ding{206}. 
Additionally, the {\tt content.js} uses the same communication channel to log cookies, storage, and APIs data on a storage.
\noindent\textbf{Data collection.}
We conduct an automated crawl of the landing pages of the top 10K websites in the Tranco top-million list \cite{LePochat2019} as of July 2023, using Selenium with Chrome 114.0.5735.133 and a purpose-built extension. 
We perform this crawl from the United States.
On average, it took approximately 10 seconds for a webpage to fully load (until the {\tt onLoad} event is fired) \cite{onloadevent}, and an additional 30 seconds before moving on to the next website. 
The crawling process is stateless, as we cleared all cookies and local browser states between consecutive crawls.
%
This helps ensure that the collected data is reproducible and accurately reflects the current state of the webpage without being biased by previous webpage visits.

\begin{figure}[t]
    \centering
      \includegraphics[width=0.23\textwidth]{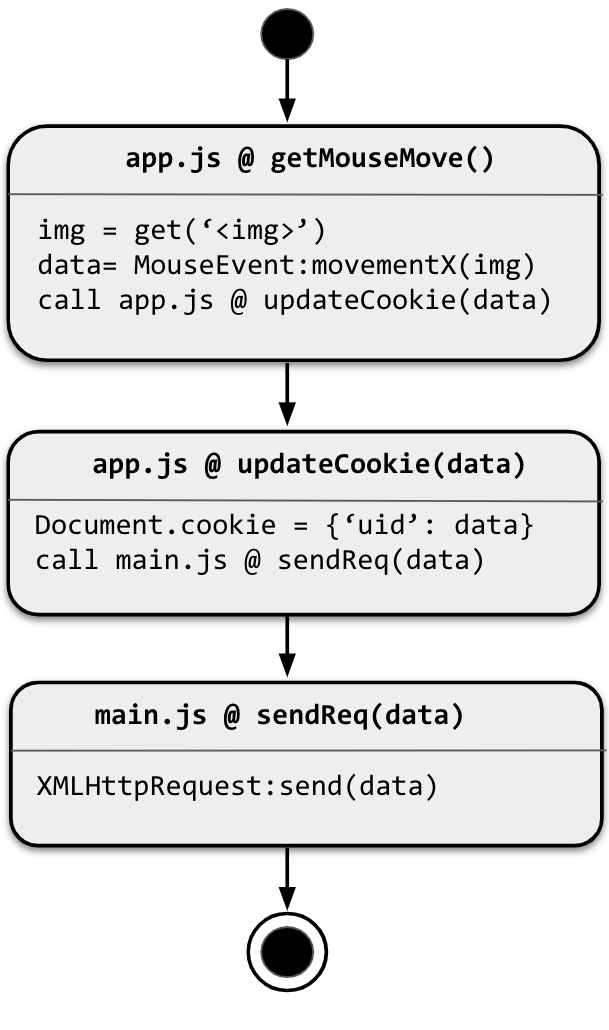}
      \vspace{-0.15in}
        \caption{{\tt getMouseMove()} collects data, calls {\tt updateCookie(data)} to modify cookie, and triggers {\tt sendReq(data)} to send the request.}
        \label{fig:scenerio}
       \vspace{-0.08in}
\end{figure}
\subsection{Graph Representation}
\tool constructs a graph representation from the collected data of each webpage.
\tool's graph representation leverages JS function-level features and JS execution context of each of the webpage’s activity, as listed in Table \ref{table:webapis}.
The graph's unique structure offers several advantages over traditional techniques. 
\tool allows for semantic reasoning with its dynamic execution context and enables traceability by providing a complete history of how a particular webpages activity is executed. 
This information is essential in differentiating the intention of the same graph node (\eg JS function) when participating in different execution scenarios. 
While their static representation remains unchanged, the JS execution context around them at runtime may alter their semantics.   
By building a graph representation first, \tool facilitates the extraction of structural and contextually rich features, which it utilizes to identify privacy-invasive JS functions that are otherwise tightly interleaved with non-tracking code.
\tool's graph representation is the first-of-its-kind \cite{Iqbal20AdgraphSP,FilterlistGeneration2020ACMWWW,Siby22WebGraph} to incorporate the JS execution call stack and calling contexts (scope chain) fully. 

\noindent\textbf{Nodes.}
\tool categorizes a webpage's activity into five types of nodes: JS functions, DOM elements, network, storage, and web APIs. 
The JS function node represents a function call that has attributes, including its parent script's URL, name (except for anonymous functions), scope chain, line, and column number. 
The scope chain refers to the variables, functions, objects, and closures that a JS function can access at the time of invocation.  
Closures are a special type of function that can access variables in its enclosing function's scope chain, even after the function has returned.
One function can generate multiple nodes depending on its calling sequence (call stack) and context (scope chain). 
This is particularly useful for gateway functions that participate in both tracking and non-tracking activity.
For the threat model in Figure \ref{fig:mixed-script-initiator}, \tool creates two separate nodes for the {\tt sendRequest} function: one for its invocation within {\tt getIdentifier} and another for {\tt loadImage}. 
The DOM element node has attributes such as element name, class, or id and a value if available. 
The network node represents all the network requests that are sent by a webpage.
The storage node represents all client-side data storing mechanisms, with attributes that differentiate between mechanisms such as cookies \cite{DocumentCookie} and local storage \cite{LocalStorage}. 
The web API node type represents Web APIs, as listed in Table \ref{table:webapis}, with attributes that differentiate between APIs, such as charging time and discharging time.
%
%
\tool's graph representation is the first to include function and web API node types.

\noindent\textbf{Edges.}
Edges in the graph generated by \tool depict runtime dependencies between nodes. 
We extract two types of edges: call and behavioral edges. 
Call edges represent the sequence of function calls in a JS execution call stack. 
These edges connect function nodes to other function nodes to represent the dynamic caller-callee relationship. 
Whenever a function is called, a directed edge is created from the caller function node to the callee function node.
These edges add valuable information about the sequence of function calls, which existing graph representations do not capture \cite{Iqbal20AdgraphSP, Siby22WebGraph}. 
Such edges enhance \tool's capabilities in modeling the semantics of the JS script by making its graph context-aware. 
On the other hand, behavioral edges are used to represent interactions between the JS functions, DOM, network, storage, and API nodes.
If a function initiates a network request, a behavioral edge is created from the corresponding JS function node to the network node. 
Similarly, if a function triggers a storage or Web API call, a behavioral edge is created between the JS function node and the corresponding storage or Web API node, respectively.
The direction of the edge is dependent on the context of the API call. 
For instance, if the JS function is storing data, there is an edge from the JS function to the storage or API node, and vice versa. 

\begin{figure}[t]
    \centering
      \includegraphics[width=0.37\textwidth]{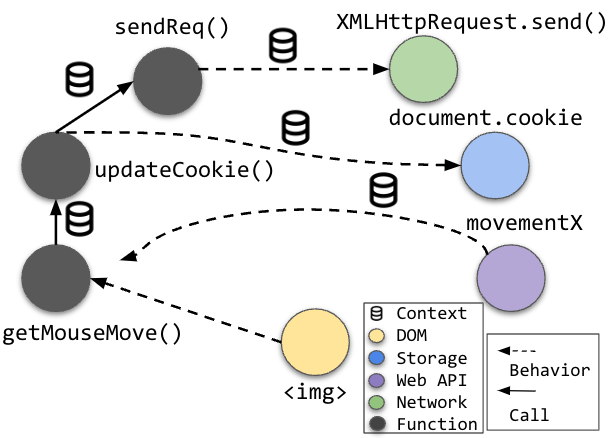}
      \put(-147,35){{\ding{202}}}
      \put(-145,65){{\ding{203}}}
      \put(-180,65){{\ding{204}}}
      \put(-130,95){{\ding{205}}}
      \put(-177,100){{\ding{206}}}
      \put(-120,115){{\ding{207}}}
      \vspace{-0.05in}
        \caption{A simplified \tool's graph on mouse movement tracking.}
        \label{fig:scenerio-graph}
    \vspace{-0.1in}
\end{figure}
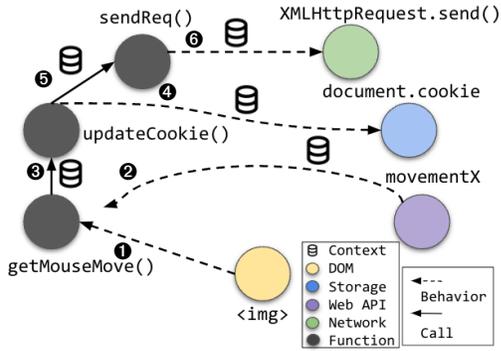
\noindent \textbf{Graph construction.}
Figure \ref{fig:scenerio} illustrates a sequence diagram for a webpage activity that tracks user movement on an HTML element, stores it in a cookie, and sends data to an external server. 
Each box in the illustration represents an entry of a JS execution call stack. 
The {\tt getMouseMove} function is responsible for capturing the movement of the mouse pointer on the specified HTML element. 
This is achieved through the {\tt MouseEvent} API, which provides access to the {\tt movementX} property.
This property represents the difference in the X coordinate of the mouse pointer between the current event and the previous mouse move event.
Once the data (context) is collected, it is passed on to the {\tt updateCookie} function, which updates the cookie value using the {\tt document.cookie} property. 
The data (context) is then passed to the {\tt sendReq} function, which sends an HTTP request to an external server with the data. 
%

Figure \ref{fig:scenerio-graph} shows a \tool's graph representation.
The graph begins with a behavioral edge (dotted-line) that obtains mouse movement data associated with the specific HTML element, represented in \ding{202} and \ding{203}. 
The call edge (solid-line) \ding{204} between {\tt getMouseMove} and {\tt updateCookie} represents a function call that also passes the mouse movement data (scope).
The behavioral edge in \ding{205} represents the {\tt document.cookie} call to update the storage node.
The call edge \ding{206} between {\tt updateCookie} and {\tt sendReq} represents a function call that also passes the mouse movement data (scope).
Finally, the behavioral edge \ding{207} between the function and the network node represents the HTTP request sent to an external server.

\subsection{Feature Extraction}
Once a fine-grained graph is generated for all observed web activities on a webpage, \tool extracts two kinds of features from the graph to augment the current node's attributes: structural and contextual-based features.

\noindent \textbf{Structural features.}
Structural features represent relationships between nodes in the graph, such as ancestry information and connectivity. 
For example, how many Web API nodes are present in the ancestor chain of a function node or whether a function directly or indirectly interacts with a storage node?
Adding JS functions from the execution call stack improves the completeness of the structural features by providing additional context under which a function is called and filling in the missing pieces about the sequence of events prior to reaching the current node.

\noindent \textbf{Contextual features.}
\tool includes JS dynamic execution context in the generated graph using contextual-based features. 
%
We extract three types of contextual features. 
First, we count the number of local variables, global variables, and closure variables. 
Second, we count the number of arguments passed to the function.
Third, we include the number of network requests, DOM modifications, and API (\eg cookies and storage APIs) calls made by the function.
These features play a vital role in understanding JS function behavior, especially during the execution of tracking activity. 
A function's calling context has been used previously to construct the dynamic invariants of a program, which helps in verifying a given program behavior~\cite{proteusFse16, NguyenIcse2014, mirshokraie2014guided}. 
Such invariants are often implemented as assertions to detect unintended program behavior. 
Similarly, \tool's contextual features can help the downstream training process learn invariants about a JS function under which the JS function participates in tracking and the invariant under which it participates in non-tracking activity.  
Our graph representation is unique in its ability to extract contextual-based features that cannot be obtained by prior approaches \cite{Iqbal20AdgraphSP,FilterlistGeneration2020ACMWWW,Siby22WebGraph}.

\begin{table}[t]
\small
    \centering
    \scalebox{0.9}{\begin{tabular}{ |l |r | r| r| }
\hline
\multirow{2}{*}{\textbf{Features}}&{\tt send}&{\tt update}&{\tt getMouse}\\
& {\tt Req()}& {\tt Cookie()}&{\tt Move()}\\
\hline
Number of requests sent &1 &1 &1\\
Is gateway function &1 &0 &0\\
Number of cookie  (setter) &0 &1 &0 \\
Number of web API  (getter)&0 &0 &1\\
Number of arguments &1 &1 &0 \\
Number of callee functions  &1 &1 &0 \\
Number of caller functions  &0 &1 &1 \\
Ascendant has cookie accesses  &1 &0 &0 \\
Descendant has cookie accesses &0 &0 &1\\
Ascendant has web API accesses  &1 &1 &0 \\

\hline
\end{tabular}
}
    \vspace{0.2in}
    \caption{A small subset of features extracted from Figure \ref{fig:scenerio-graph}.}
    \label{table:feature}
   \vspace{-0.30in} 
\end{table}
Table \ref{table:feature} presents a subset of the features, including structural and contextual attributes, obtained from the graph representation of \tool shown in Figure \ref{fig:scenerio-graph}. 
The {\tt number of requests sent} feature is triggered for all functions, as the function calls appear in the call stack when tracking requests are initiated, with the {\tt sendReq} function acting as the gateway. 
Contextual features encompass the number of storage (getter and setter) and web API (getter and setter) operations, as well as function arguments, providing insights into each function's contextual behavior. 
Additionally, including structural attributes, specifically ascendant and descendant relationships in storage and web API nodes, enhances comprehension of hierarchical structures and code dependencies
For a detailed catalog of all features, please refer to Table \ref{table:Allfeatures} in the appendix.

\subsection{Labeling} 
We adopt a prior approach \cite{amjad2023blocking} to label \tool's graph representation.
We label network requests as either tracking or non-tracking based on whether their URL matches the rules in EasyList \cite{easylist} and EasyPrivacy \cite{easyprivacy}, which are widely employed by content blockers to identify tracking activity. 
While the filter lists do have false negatives, they are highly accurate---network request labeled as tracking by filter lists is actually tracking---as each rule in the filter list is vetted manually by experts. 
Next, we analyze the call stack of the requests and label a JS function as tracking if it exclusively participates in the stack trace of tracking requests. 
Otherwise, if a JS function participates in non-tracking requests (or a combination of tracking and non-tracking requests), we label it as a non-tracking JS function. 
This approach establishes a conservative ground truth where the functions with mixed behavior are labeled as non-tracking.
However, these mixed functions comprise only 3.9\% of our ground truth, as discussed in more detail in Section \ref{sec:limitations}.
In addition, we exclude functions that trigger web, storage, or cookie API calls but are not present in either a tracking or non-tracking request call stack. 
This is because we lack evidence from filter lists to label them.

\subsection{Classification}
We use a random forest classifier, which is an ensemble learning method. 
It constructs multiple decision trees and combines their predictions to obtain a final prediction. 
Prior work \cite{Iqbal20AdgraphSP, Siby22WebGraph} also use a random forest classifier since it outperformed other comparable models.
To assess our classifier's performance, we divide our dataset into training, validation, and testing sets.
Before splitting, we deduplicate our dataset by considering only one instance for each script URL and function name, ensuring no contamination between the training and testing phases. 
The numbers in Table \ref{table:databreakdown} represent unique functions.
Our dataset initially contains 3.0 million total functions, which reduce to 1.8 million unique functions after deduplication.
Specifically, we use 60\% of the dataset for training the classifier, 20\% for evaluating its accuracy during hyperparameter tuning, and the remaining 20\% for final evaluation of the model. 
To optimize hyperparameters, we use the validation set to configure the depth of each decision tree in the forest to 20 and the number of trees used in the forest to 1000.

\begin{table}[t]
\small
    \centering
    \scalebox{1}{\begin{tabular}{ |l| r | r| r|}
\hline
\textbf{Data-split}& \textbf{Tracking} &\textbf{Non-tracking}&\textbf{Total}\\
& \textbf{Functions} &\textbf{Functions}&\textbf{Functions}\\
\hline
Training & 408,429 & 674,724 & 1,083,153 \\
Validation & 135,590 & 225,462 & 361,052 \\
Testing & 136,045 & 225,007 & 361,052 \\
\hline
\end{tabular}} 
    \vspace{0.2in}
    \caption{The breakdown of the data employed to train, test, and validate the \tool classifier.}
    \label{table:databreakdown}
    \vspace{-0.3in}
\end{table}

\subsection{Surrogate Generation and Replacement}
%
\tool generates surrogate scripts to replace mixed scripts in future page loads. 
\tool's surrogate generation and replacement is website-specific, 
which helps address the variations from obfuscation and minification techniques deployed by different websites. 
%
Our approach for surrogate generation involves neutralizing tracking function calls by substituting them with a mock function call within the script. 
Once surrogate scripts are generated for a specific website, we create a filter rule that substitutes the original script with the surrogate script at runtime. 

\noindent\textbf{Surrogate generation.}
Our surrogate script generation process relies on three key elements: (1) the classification labels assigned by \tool, (2) the script source as it appeared when the response was received, and (3) the line and column numbers corresponding to the function call at runtime. 
\begin{algorithm}
\small
\caption{Response Replacement Algorithm for Chrome Extension Manifest version 2.
\vspace{0.1in}}
\label{algo:ResponseReplacementAlgo}
\SetAlgoNlRelativeSize{-3}
\SetNlSty{}{}{:}
\KwIn{$scriptURL$ --- the script URL to be requested}
\KwOut{Returns $surrogate$ if available, otherwise $None$ to indicate presence or absence of a replacement.}
\SetKwFunction{checkFilter}{isInNOTJSFilter}
\SetKwFunction{getResponse}{getResponse}
\SetKwFunction{getSurrogate}{getSurrogateResponse}
\SetKwFunction{continueReq}{Fetch.continueRequest}

\SetKwProg{Pn}{Procedure}{}{}
\Pn{{$message, URL, responseCode$}}{
    \If{$message == \text{"Fetch.requestPaused"}$}{
        $response \gets \getResponse{}$\;
        \If{$responseCode == 200$}{
            \If{\checkFilter{$URL$}}{
                $surrogate \gets \getSurrogate{}$\;
                \If{$surrogate \neq None$}{
                    $response \gets surrogate$\;
                }
            }
        }
    }
    \continueReq{$response$}\;}
\end{algorithm}

Leveraging this information, we neutralize tracking function calls by substituting them with a mock call designed to consistently return an empty response.
It is important to note that surrogate generation is an offline step.
To illustrate this process, consider the function call {\tt track(o[0], o[1])} on {\tt adobe.com} in Listing \ref{lst:surrogate}
that operates in the context of Adobe Analytics 
and is classified as tracking by \tool. 
%
%
\tool records the exact line (line 7) and column where this function call begins. The column corresponds to the first element {\tt "t"} of the function name  {\tt "track"} in this example.
In the first step, we verify that the function call exists at the recorded line and column numbers based on its function name while skipping this step for anonymous functions.
We replace the original function call with the mock call that returns an empty response.
As we discuss later in Section \ref{subsec: surrogate generation}, this simple approach effectively neutralizes a vast majority of tracking function calls while upholding the syntactic integrity of the script. 

\lstdefinestyle{base}{
language=Java,
moredelim=**[is][\color{red}]{@}{@},
moredelim=**[is][\color{darkgreen}]{~}{~},
numbers=left,
numberstyle=\tiny,
stepnumber=1,
numbersep=5pt,
escapechar=`
}

\begin{lstlisting}[caption={The example demonstrating the neutralization of tracking function calls during surrogate generation.},language=Java, label={lst:surrogate}, style=base]
function x() {
    var e = [];
    ..., 
    t.__satelliteLoadedCallback((function() {
        var n, a, o;
        for (n = 0, a = e.length; n < a; n++) o = e[n], 
        `\addtocounter{lstnumber}{-1}`@-t._satellite.track(o[0], o[1])@
        ~+t._satellite.mockTrack()~ 
    })), _satellite.track("pageload")}
\end{lstlisting}

\noindent\textbf{Surrogate replacement.}
\tool replaces the original mixed scripts with the generated surrogates at runtime during future page loads.
%
\tool first identifies the target scripts on a webpage using regular expressions (regex) of generated surrogate scripts.
For example, we create a regex rule to identify scripts associated with a domain like {\tt adobe.com/*analytics.js}. 
Once identified, \tool replaces the target script with the corresponding surrogate. 
This replacement mechanism is supported in different Chrome extension environments like manifest versions 2 (V2) and 3 (V3).
%
%
In manifest V2, Chrome extensions employ the Fetch API \cite{FetchAPI}, a conventional method for intercepting and modifying responses at runtime. 
Algorithm \ref{algo:ResponseReplacementAlgo} shows the steps for surrogate replacement.
When a network request is made, the \tool's browser extension intercepts it, verifies that the response status is {\tt OK} (status code 200), and subsequently modifies the response content. 
This approach allows us to replace a mixed script with the corresponding surrogate script. 
It is worth noting that this approach is similar to scriptlet replacement in content blocking tools such as uBlock Origin \cite{ubo-scriptlets} and AdGuard \cite{adguardsnippets}.
Manifest V3 introduces the Declarative Net Request API \cite{NETRequestAPI} that does not allow direct response modification capabilities like V2.
Instead, the original request is blocked and redirected to an alternate URL in manifest V3, effectively replacing the response content with the content retrieved from the redirected URL ~\cite{ManifestV3}.
%

\begin{table}[t]
\small
    \centering
    \scalebox{1}{\begin{tabular}{ |l| l|c|c| c|}
\hline
\textbf{Model} & \textbf{Section}&\textbf{Precision} & \textbf{Recall} &\textbf{F1}\\
& & & &\textbf{Score}\\
\hline
    \tool & Standard - 5.1 & 94.3\% & 98.0\% & 96.2\%\\
     \tool  & Obfuscation - 5.3 & 93.5\% & 90.4\% & 91.9\% \\
    \tool  & Coverage - 5.3 & 88.4\% & 95.7\% & 91.9\% \\
    \webgr& Comparison - 5.4 &49.3\% & 66.4\% & 56.5\%\\
    SugarCoat & Comparison - 5.4 & 23.0\% & 22.6\% & 22.8\%\\
\hline
\end{tabular}}
    \vspace{0.2in}
    \caption{\tool's precision, recall, F1-score in standard settings, enhanced coverage, obfuscation robustness, and comparison with existing tools.}
    \label{table:robustness}
    \vspace{-0.27in}
\end{table}

\section{Evaluation}
\label{sec:eval}
This section evaluates \tool's accuracy, feature contributions,  robustness to JS obfuscation, and enhanced code coverage and compares it to existing countermeasures. 
It also evaluates  \tool's automated surrogate generation, replacement, and user-centric manual breakage inspection.

\subsection{Accuracy Analysis}
We train \tool's random forest classifier on approximately 1.1 million JS functions in the training set, do hyper-parameter tuning on 361 thousand JS functions in the validation set, and then evaluate its accuracy on 361 thousand JS functions in the testing set.
Table \ref{table:databreakdown} presents a breakdown of the data split between tracking and non-tracking functions.
\tool is able to achieve a precision of 94.3\% and a recall of 98.0\%. 
The overall F1-score for \tool is 96.2\%, indicating its effectiveness in accurately distinguishing between tracking and non-tracking JS functions.
A confusion matrix (Table~\ref{table:confusionmatrix}) of this analysis is attached in the appendix.
Additionally, we perform 5-fold cross-validation to assess \tool's accuracy. 
We split the dataset into five equally sized folds, where the model is trained on four folds and evaluated on the remaining fold, repeated five times so that each fold serves as the testing set once.
In cross-validation, \tool achieves a comparable precision of 94.6 and a recall of 96.2, with an overall 95.3$\pm$0.3\% F1-score, averaged over 5-folds.

\noindent \textbf{Error analysis.}
While \tool has a relatively low false positive rate (3.5\%) and false negative rate (1.9\%), we investigate the reasons and contexts of its errors.
%
We conduct a manual evaluation by randomly sampling 50 instances where \tool incorrectly identifies JS functions as tracking, despite our ground-truth data missing them. 
The causes for these errors can be categorized into three primary categories: 
\begin{table}[t]
\small
    \centering
    \scalebox{0.98}{\begin{tabular}{ |l |c | r|}
\hline
\textbf{Feature}& \textbf{Type} &\textbf{Information Gain (\%)}\\
\hline
Number of successor functions & Structural & 15.1\% \\
Number of local storage accesses & Contextual  & 11.9\% \\
Number of descendant nodes & Structural & 9.8\% \\
Number of in-edges & Structural & 7.8\% \\
Number of Web API calls & Contextual & 4.8\% \\
\hline

\end{tabular}}
    \vspace{0.15in}
    \caption{Top five features that contribute the most to \tool’s accuracy.}
    \label{table:information-gain}
    \vspace{-0.3in}
\end{table}

The first category includes functions in mixed scripts that cannot be blocked by filter lists to prevent website functionality breakage.
For example, the function {\tt t} in {\tt js.cookie.min.js} on {\tt kakaku.com} sets the tracking cookie and is classified as tracking by \tool, while its other functions primarily serve the history feature on the website.
Consequently, being a mixed script, it cannot be blocked by filter list authors (acknowledged in this GitHub issue \cite{18242}), yet it can be handled by \tool.

The second category includes functions that are actually involved in tracking but missed by filter lists \cite{alrizah2019errors}.
For example, the function {\tt b} in the script {\tt htlbid-advertising.min.js} on {\tt wkrn.com} manages ad slots and their configurations. 
We conduct an in-depth manual evaluation of the entire script, discovering that most of its functions are classified as tracking by \tool.
Following this, we report this issue \cite{18230} to filter list authors, leading to its inclusion in the filter rules.
In total, \tool identifies ten such cases out of a sample of fifty in the aforementioned two categories that are either missed or cannot be handled by EasyList/EasyPrivacy.
We report these cases to filter list authors, leading to four \cite{18230,18242,18185, 18243} being recognized and six still pending review, highlighting \tool's superior detection capabilities compared to EasyList/EasyPrivacy.

The last category includes functions, where forty out of fifty instances represent an actual error by \tool. 
We use the number of arguments, local and global variables in the scope chain, rather than examining the types or values passed to these variables. 
Therefore, functions with identical dynamic execution contexts—\ie the same call stack and scope chain—for both tracking and non-tracking activities are labeled as non-tracking in the ground truth but classified as tracking by \tool. 
These instances are not technically misclassified by \tool, but rather point to the limitations intrinsic to dynamic execution context.
Such multi-purpose functions can serve both tracking and non-tracking roles. 
For instance, in Listing \ref{lst:mixedfunction}, {\tt \_setField} function in {\tt visitor*.js} script served by {\tt microsoft.com}.
If parameter {\tt f} is set to false, the function participates in the tracking activity on the website.
However, if parameter {\tt f} is set to true, the function participates in the non-tracking activity on the website.
In ground truth, this function is labeled as non-tracking because of the same number of arguments. 
A deeper examination of parameter types and values could address these issues in future models.
In summary, 20\% of the tracking functions in the sample of fifty represent errors in the filter list, while the remaining 80\% constitute actual errors by \tool.

\begin{lstlisting}[language=Java, label={lst:mixedfunction},caption={\tt \_setField} function from the {\tt Microsoft} script domain]
a._setField = function(b, d, f) {
        null == a._fields && (a._fields = {});
        a._fields[b] = d;
        f || a._writeVisitor()};
\end{lstlisting}

%
False negatives predominantly stem from code coverage, as not all properties of tracking functions are captured.
For instance, a more comprehensive list of Web APIs could be employed to capture additional characteristics, which would also assist in more precise profiling of tracking functions and, consequently, fewer false negatives.
More comprehensive crawling can help address these issues, but it may lead to more noise and higher graph construction costs.  

\subsection{Feature Analysis}
\begin{figure}[!t]
    \begin{minipage}[t]{0.48\textwidth}
        \begin{tikzpicture}
\begin{axis}[
    ylabel={CDF},
    legend pos=south east,
    ymin=0,
    ymax=1,
    xmin=0,
    xmax=2000,
    grid=both,
    line width=2pt,
    axis line style={thick},
    width=0.9*\textwidth,
    height=4.2cm,
]

\addplot[color=green] coordinates {
(-54.023, 0.000)
(-40.390, 0.004)
(-26.756, 0.030)
(-13.123, 0.122)
(0.510, 0.296)
(14.144, 0.482)
(27.777, 0.599)
(41.411, 0.650)
(55.044, 0.673)
(68.677, 0.688)
(82.311, 0.701)
(95.944, 0.713)
(109.578, 0.723)
(123.211, 0.733)
(136.844, 0.742)
(150.478, 0.751)
(164.111, 0.758)
(177.745, 0.766)
(191.378, 0.773)
(205.011, 0.780)
(218.645, 0.787)
(232.278, 0.794)
(245.912, 0.801)
(259.545, 0.807)
(273.178, 0.814)
(286.812, 0.820)
(300.445, 0.825)
(314.079, 0.831)
(327.712, 0.837)
(341.345, 0.843)
(354.979, 0.849)
(368.612, 0.855)
(382.246, 0.860)
(395.879, 0.864)
(409.512, 0.868)
(423.146, 0.872)
(436.779, 0.875)
(450.413, 0.879)
(464.046, 0.882)
(477.679, 0.886)
(491.313, 0.889)
(504.946, 0.893)
(518.580, 0.896)
(532.213, 0.899)
(545.846, 0.902)
(559.480, 0.905)
(573.113, 0.908)
(586.747, 0.911)
(600.380, 0.914)
(614.013, 0.917)
(627.647, 0.920)
(641.280, 0.923)
(654.914, 0.925)
(668.547, 0.928)
(682.180, 0.930)
(695.814, 0.932)
(709.447, 0.934)
(723.081, 0.936)
(736.714, 0.938)
(750.347, 0.940)
(763.981, 0.942)
(777.614, 0.945)
(791.248, 0.946)
(804.881, 0.948)
(818.514, 0.950)
(832.148, 0.951)
(845.781, 0.952)
(859.415, 0.954)
(873.048, 0.955)
(886.681, 0.957)
(900.315, 0.958)
(913.948, 0.959)
(927.582, 0.961)
(941.215, 0.962)
(954.848, 0.963)
(968.482, 0.965)
(982.115, 0.966)
(995.749, 0.967)
(1009.382, 0.968)
(1023.015, 0.969)
(1036.649, 0.970)
(1050.282, 0.972)
(1063.916, 0.973)
(1077.549, 0.974)
(1091.182, 0.975)
(1104.816, 0.976)
(1118.449, 0.977)
(1132.083, 0.978)
(1145.716, 0.979)
(1159.349, 0.980)
(1172.983, 0.980)
(1186.616, 0.981)
(1200.250, 0.982)
(1213.883, 0.982)
(1227.516, 0.983)
(1241.150, 0.984)
(1254.783, 0.984)
(1268.417, 0.985)
(1282.050, 0.985)
(1295.683, 0.986)
(1309.317, 0.986)
(1322.950, 0.987)
(1336.583, 0.987)
(1350.217, 0.988)
(1363.850, 0.988)
(1377.484, 0.989)
(1391.117, 0.989)
(1404.750, 0.989)
(1418.384, 0.990)
(1432.017, 0.990)
(1445.651, 0.991)
(1459.284, 0.991)
(1472.917, 0.992)
(1486.551, 0.992)
(1500.184, 0.993)
(1513.818, 0.993)
(1527.451, 0.993)
(1541.084, 0.994)
(1554.718, 0.994)
(1568.351, 0.994)
(1581.985, 0.994)
(1595.618, 0.994)
(1609.251, 0.995)
(1622.885, 0.995)
(1636.518, 0.995)
(1650.152, 0.995)
(1663.785, 0.995)
(1677.418, 0.996)
(1691.052, 0.996)
(1704.685, 0.996)
(1718.319, 0.997)
(1731.952, 0.997)
(1745.585, 0.997)
(1759.219, 0.997)
(1772.852, 0.997)
(1786.486, 0.997)
(1800.119, 0.997)
(1813.752, 0.998)
(1827.386, 0.998)
(1841.019, 0.998)
(1854.653, 0.998)
(1868.286, 0.998)
(1881.919, 0.998)
(1895.553, 0.998)
(1909.186, 0.998)
(1922.820, 0.998)
(1936.453, 0.998)
(1950.086, 0.998)
(1963.720, 0.998)
(1977.353, 0.998)
(1990.987, 0.999)
(2004.620, 0.999)
(2018.253, 0.999)
(2031.887, 0.999)
(2045.520, 0.999)
(2059.154, 0.999)
(2072.787, 0.999)
(2086.420, 0.999)
(2100.054, 0.999)
(2113.687, 0.999)
(2127.321, 0.999)
(2140.954, 0.999)
(2154.587, 0.999)
(2168.221, 0.999)
(2181.854, 0.999)
(2195.488, 0.999)
(2209.121, 0.999)
(2222.754, 0.999)
(2236.388, 0.999)
(2250.021, 0.999)
(2263.655, 0.999)
(2277.288, 0.999)
(2290.921, 0.999)
(2304.555, 0.999)
(2318.188, 0.999)
(2331.822, 0.999)
(2345.455, 0.999)
(2359.088, 0.999)
(2372.722, 0.999)
(2386.355, 0.999)
(2399.989, 0.999)
(2413.622, 0.999)
(2427.255, 0.999)
(2440.889, 0.999)
(2454.522, 0.999)
(2468.156, 0.999)
(2481.789, 0.999)
(2495.422, 1.000)
(2509.056, 1.000)
(2522.689, 1.000)
(2536.323, 1.000)
(2549.956, 1.000)
(2563.589, 1.000)
(2577.223, 1.000)
(2590.856, 1.000)
(2604.490, 1.000)
(2618.123, 1.000)
(2631.756, 1.000)
(2645.390, 1.000)
(2659.023, 1.000)
};
\addplot[color=red, dashed] coordinates {
(-54.023, 0.000)
(-40.390, 0.011)
(-26.756, 0.038)
(-13.123, 0.093)
(0.510, 0.179)
(14.144, 0.285)
(27.777, 0.389)
(41.411, 0.472)
(55.044, 0.530)
(68.677, 0.567)
(82.311, 0.590)
(95.944, 0.607)
(109.578, 0.620)
(123.211, 0.631)
(136.844, 0.641)
(150.478, 0.649)
(164.111, 0.658)
(177.745, 0.666)
(191.378, 0.674)
(205.011, 0.682)
(218.645, 0.689)
(232.278, 0.696)
(245.912, 0.703)
(259.545, 0.709)
(273.178, 0.715)
(286.812, 0.721)
(300.445, 0.727)
(314.079, 0.733)
(327.712, 0.739)
(341.345, 0.745)
(354.979, 0.751)
(368.612, 0.758)
(382.246, 0.764)
(395.879, 0.770)
(409.512, 0.776)
(423.146, 0.781)
(436.779, 0.787)
(450.413, 0.792)
(464.046, 0.797)
(477.679, 0.802)
(491.313, 0.806)
(504.946, 0.810)
(518.580, 0.814)
(532.213, 0.819)
(545.846, 0.823)
(559.480, 0.827)
(573.113, 0.831)
(586.747, 0.835)
(600.380, 0.839)
(614.013, 0.843)
(627.647, 0.847)
(641.280, 0.850)
(654.914, 0.854)
(668.547, 0.857)
(682.180, 0.860)
(695.814, 0.863)
(709.447, 0.866)
(723.081, 0.869)
(736.714, 0.872)
(750.347, 0.875)
(763.981, 0.879)
(777.614, 0.882)
(791.248, 0.885)
(804.881, 0.888)
(818.514, 0.890)
(832.148, 0.893)
(845.781, 0.896)
(859.415, 0.898)
(873.048, 0.901)
(886.681, 0.904)
(900.315, 0.906)
(913.948, 0.908)
(927.582, 0.911)
(941.215, 0.913)
(954.848, 0.915)
(968.482, 0.917)
(982.115, 0.919)
(995.749, 0.921)
(1009.382, 0.923)
(1023.015, 0.925)
(1036.649, 0.927)
(1050.282, 0.929)
(1063.916, 0.930)
(1077.549, 0.932)
(1091.182, 0.934)
(1104.816, 0.936)
(1118.449, 0.938)
(1132.083, 0.940)
(1145.716, 0.942)
(1159.349, 0.944)
(1172.983, 0.945)
(1186.616, 0.946)
(1200.250, 0.947)
(1213.883, 0.949)
(1227.516, 0.950)
(1241.150, 0.952)
(1254.783, 0.953)
(1268.417, 0.955)
(1282.050, 0.956)
(1295.683, 0.957)
(1309.317, 0.958)
(1322.950, 0.959)
(1336.583, 0.960)
(1350.217, 0.962)
(1363.850, 0.963)
(1377.484, 0.964)
(1391.117, 0.965)
(1404.750, 0.966)
(1418.384, 0.966)
(1432.017, 0.967)
(1445.651, 0.968)
(1459.284, 0.969)
(1472.917, 0.970)
(1486.551, 0.972)
(1500.184, 0.973)
(1513.818, 0.974)
(1527.451, 0.974)
(1541.084, 0.975)
(1554.718, 0.976)
(1568.351, 0.977)
(1581.985, 0.977)
(1595.618, 0.978)
(1609.251, 0.978)
(1622.885, 0.979)
(1636.518, 0.979)
(1650.152, 0.980)
(1663.785, 0.980)
(1677.418, 0.981)
(1691.052, 0.982)
(1704.685, 0.982)
(1718.319, 0.983)
(1731.952, 0.984)
(1745.585, 0.984)
(1759.219, 0.984)
(1772.852, 0.985)
(1786.486, 0.985)
(1800.119, 0.985)
(1813.752, 0.986)
(1827.386, 0.986)
(1841.019, 0.986)
(1854.653, 0.986)
(1868.286, 0.986)
(1881.919, 0.986)
(1895.553, 0.987)
(1909.186, 0.987)
(1922.820, 0.987)
(1936.453, 0.987)
(1950.086, 0.988)
(1963.720, 0.988)
(1977.353, 0.989)
(1990.987, 0.989)
(2004.620, 0.990)
(2018.253, 0.990)
(2031.887, 0.990)
(2045.520, 0.990)
(2059.154, 0.990)
(2072.787, 0.990)
(2086.420, 0.990)
(2100.054, 0.990)
(2113.687, 0.990)
(2127.321, 0.990)
(2140.954, 0.990)
(2154.587, 0.990)
(2168.221, 0.990)
(2181.854, 0.990)
(2195.488, 0.991)
(2209.121, 0.991)
(2222.754, 0.991)
(2236.388, 0.991)
(2250.021, 0.992)
(2263.655, 0.992)
(2277.288, 0.992)
(2290.921, 0.992)
(2304.555, 0.992)
(2318.188, 0.992)
(2331.822, 0.992)
(2345.455, 0.992)
(2359.088, 0.993)
(2372.722, 0.993)
(2386.355, 0.994)
(2399.989, 0.994)
(2413.622, 0.994)
(2427.255, 0.995)
(2440.889, 0.995)
(2454.522, 0.995)
(2468.156, 0.995)
(2481.789, 0.995)
(2495.422, 0.996)
(2509.056, 0.996)
(2522.689, 0.996)
(2536.323, 0.996)
(2549.956, 0.996)
(2563.589, 0.996)
(2577.223, 0.996)
(2590.856, 0.996)
(2604.490, 0.996)
(2618.123, 0.996)
(2631.756, 0.996)
(2645.390, 0.996)
(2659.023, 0.996)
};

\legend{Non-tracking, Tracking}

\end{axis}
\end{tikzpicture}
        \subcaption{Number of successor functions}
        \label{fig:NumOfMethSucces}
        \vspace{0.1in}
    \end{minipage}
    \begin{minipage}[t]{0.48\textwidth}
        \begin{tikzpicture}
\begin{axis}[
    ylabel={CDF},
    legend pos=south east,
    ymin=0,
    ymax=1,
    xmin=0,
    xmax=200,
    grid=both,
    line width=2pt,
    axis line style={thick},
    width=0.9*\textwidth,
    height=4.2cm,
]

\addplot[color=green] coordinates {
(-3.728, 0.000)
(-2.333, 0.006)
(-0.939, 0.101)
(0.455, 0.400)
(1.849, 0.612)
(3.244, 0.669)
(4.638, 0.700)
(6.032, 0.727)
(7.426, 0.748)
(8.821, 0.765)
(10.215, 0.780)
(11.609, 0.794)
(13.003, 0.807)
(14.398, 0.819)
(15.792, 0.830)
(17.186, 0.840)
(18.580, 0.850)
(19.975, 0.860)
(21.369, 0.869)
(22.763, 0.877)
(24.157, 0.885)
(25.552, 0.892)
(26.946, 0.898)
(28.340, 0.905)
(29.734, 0.911)
(31.129, 0.917)
(32.523, 0.923)
(33.917, 0.928)
(35.311, 0.933)
(36.706, 0.937)
(38.100, 0.941)
(39.494, 0.944)
(40.888, 0.948)
(42.283, 0.951)
(43.677, 0.954)
(45.071, 0.957)
(46.465, 0.960)
(47.860, 0.962)
(49.254, 0.965)
(50.648, 0.967)
(52.042, 0.968)
(53.437, 0.970)
(54.831, 0.972)
(56.225, 0.974)
(57.619, 0.975)
(59.014, 0.977)
(60.408, 0.979)
(61.802, 0.980)
(63.196, 0.981)
(64.591, 0.982)
(65.985, 0.983)
(67.379, 0.984)
(68.773, 0.984)
(70.168, 0.985)
(71.562, 0.985)
(72.956, 0.986)
(74.350, 0.987)
(75.745, 0.987)
(77.139, 0.988)
(78.533, 0.989)
(79.927, 0.989)
(81.321, 0.990)
(82.716, 0.990)
(84.110, 0.990)
(85.504, 0.990)
(86.898, 0.991)
(88.293, 0.991)
(89.687, 0.991)
(91.081, 0.991)
(92.475, 0.991)
(93.870, 0.992)
(95.264, 0.992)
(96.658, 0.992)
(98.052, 0.992)
(99.447, 0.992)
(100.841, 0.993)
(102.235, 0.993)
(103.629, 0.993)
(105.024, 0.993)
(106.418, 0.993)
(107.812, 0.993)
(109.206, 0.994)
(110.601, 0.994)
(111.995, 0.994)
(113.389, 0.995)
(114.783, 0.995)
(116.178, 0.995)
(117.572, 0.996)
(118.966, 0.996)
(120.360, 0.996)
(121.755, 0.996)
(123.149, 0.997)
(124.543, 0.997)
(125.937, 0.997)
(127.332, 0.997)
(128.726, 0.997)
(130.120, 0.997)
(131.514, 0.997)
(132.909, 0.997)
(134.303, 0.997)
(135.697, 0.997)
(137.091, 0.998)
(138.486, 0.998)
(139.880, 0.998)
(141.274, 0.998)
(142.668, 0.998)
(144.063, 0.998)
(145.457, 0.998)
(146.851, 0.998)
(148.245, 0.998)
(149.640, 0.998)
(151.034, 0.998)
(152.428, 0.998)
(153.822, 0.998)
(155.217, 0.998)
(156.611, 0.998)
(158.005, 0.998)
(159.399, 0.999)
(160.794, 0.999)
(162.188, 0.999)
(163.582, 0.999)
(164.976, 0.999)
(166.371, 0.999)
(167.765, 0.999)
(169.159, 0.999)
(170.553, 0.999)
(171.948, 0.999)
(173.342, 0.999)
(174.736, 0.999)
(176.130, 0.999)
(177.525, 0.999)
(178.919, 0.999)
(180.313, 0.999)
(181.707, 0.999)
(183.102, 0.999)
(184.496, 0.999)
(185.890, 0.999)
(187.284, 0.999)
(188.679, 0.999)
(190.073, 0.999)
(191.467, 0.999)
(192.861, 0.999)
(194.255, 0.999)
(195.650, 0.999)
(197.044, 0.999)
(198.438, 0.999)
(199.832, 0.999)
(201.227, 0.999)
(202.621, 0.999)
(204.015, 0.999)
(205.409, 0.999)
(206.804, 0.999)
(208.198, 0.999)
(209.592, 0.999)
(210.986, 0.999)
(212.381, 0.999)
(213.775, 1.000)
(215.169, 1.000)
(216.563, 1.000)
(217.958, 1.000)
(219.352, 1.000)
(220.746, 1.000)
(222.140, 1.000)
(223.535, 1.000)
(224.929, 1.000)
(226.323, 1.000)
(227.717, 1.000)
(229.112, 1.000)
(230.506, 1.000)
(231.900, 1.000)
(233.294, 1.000)
(234.689, 1.000)
(236.083, 1.000)
(237.477, 1.000)
(238.871, 1.000)
(240.266, 1.000)
(241.660, 1.000)
(243.054, 1.000)
(244.448, 1.000)
(245.843, 1.000)
(247.237, 1.000)
(248.631, 1.000)
(250.025, 1.000)
(251.420, 1.000)
(252.814, 1.000)
(254.208, 1.000)
(255.602, 1.000)
(256.997, 1.000)
(258.391, 1.000)
(259.785, 1.000)
(261.179, 1.000)
(262.574, 1.000)
(263.968, 1.000)
(265.362, 1.000)
(266.756, 1.000)
(268.151, 1.000)
(269.545, 1.000)
(270.939, 1.000)
(272.333, 1.000)
(273.728, 1.000)
};
\addplot[color=red, dashed] coordinates {
(-3.728, 0.000)
(-2.333, 0.043)
(-0.939, 0.149)
(0.455, 0.307)
(1.849, 0.457)
(3.244, 0.552)
(4.638, 0.601)
(6.032, 0.628)
(7.426, 0.648)
(8.821, 0.665)
(10.215, 0.680)
(11.609, 0.693)
(13.003, 0.706)
(14.398, 0.717)
(15.792, 0.728)
(17.186, 0.739)
(18.580, 0.749)
(19.975, 0.758)
(21.369, 0.768)
(22.763, 0.777)
(24.157, 0.786)
(25.552, 0.795)
(26.946, 0.804)
(28.340, 0.812)
(29.734, 0.820)
(31.129, 0.828)
(32.523, 0.835)
(33.917, 0.842)
(35.311, 0.849)
(36.706, 0.855)
(38.100, 0.860)
(39.494, 0.865)
(40.888, 0.870)
(42.283, 0.875)
(43.677, 0.879)
(45.071, 0.883)
(46.465, 0.887)
(47.860, 0.892)
(49.254, 0.896)
(50.648, 0.899)
(52.042, 0.903)
(53.437, 0.906)
(54.831, 0.910)
(56.225, 0.913)
(57.619, 0.915)
(59.014, 0.918)
(60.408, 0.920)
(61.802, 0.923)
(63.196, 0.925)
(64.591, 0.927)
(65.985, 0.928)
(67.379, 0.929)
(68.773, 0.930)
(70.168, 0.931)
(71.562, 0.932)
(72.956, 0.934)
(74.350, 0.935)
(75.745, 0.936)
(77.139, 0.938)
(78.533, 0.939)
(79.927, 0.940)
(81.321, 0.941)
(82.716, 0.941)
(84.110, 0.942)
(85.504, 0.942)
(86.898, 0.943)
(88.293, 0.944)
(89.687, 0.945)
(91.081, 0.945)
(92.475, 0.946)
(93.870, 0.947)
(95.264, 0.948)
(96.658, 0.949)
(98.052, 0.949)
(99.447, 0.950)
(100.841, 0.951)
(102.235, 0.952)
(103.629, 0.952)
(105.024, 0.953)
(106.418, 0.954)
(107.812, 0.955)
(109.206, 0.956)
(110.601, 0.957)
(111.995, 0.959)
(113.389, 0.960)
(114.783, 0.962)
(116.178, 0.963)
(117.572, 0.964)
(118.966, 0.965)
(120.360, 0.966)
(121.755, 0.966)
(123.149, 0.967)
(124.543, 0.967)
(125.937, 0.967)
(127.332, 0.968)
(128.726, 0.968)
(130.120, 0.968)
(131.514, 0.969)
(132.909, 0.970)
(134.303, 0.971)
(135.697, 0.971)
(137.091, 0.972)
(138.486, 0.973)
(139.880, 0.973)
(141.274, 0.973)
(142.668, 0.973)
(144.063, 0.974)
(145.457, 0.974)
(146.851, 0.975)
(148.245, 0.975)
(149.640, 0.975)
(151.034, 0.976)
(152.428, 0.976)
(153.822, 0.976)
(155.217, 0.976)
(156.611, 0.977)
(158.005, 0.977)
(159.399, 0.977)
(160.794, 0.978)
(162.188, 0.979)
(163.582, 0.979)
(164.976, 0.980)
(166.371, 0.980)
(167.765, 0.980)
(169.159, 0.980)
(170.553, 0.980)
(171.948, 0.980)
(173.342, 0.980)
(174.736, 0.980)
(176.130, 0.980)
(177.525, 0.981)
(178.919, 0.981)
(180.313, 0.981)
(181.707, 0.981)
(183.102, 0.981)
(184.496, 0.981)
(185.890, 0.981)
(187.284, 0.981)
(188.679, 0.982)
(190.073, 0.982)
(191.467, 0.982)
(192.861, 0.982)
(194.255, 0.982)
(195.650, 0.983)
(197.044, 0.983)
(198.438, 0.983)
(199.832, 0.983)
(201.227, 0.983)
(202.621, 0.983)
(204.015, 0.983)
(205.409, 0.983)
(206.804, 0.984)
(208.198, 0.984)
(209.592, 0.984)
(210.986, 0.984)
(212.381, 0.984)
(213.775, 0.984)
(215.169, 0.985)
(216.563, 0.985)
(217.958, 0.985)
(219.352, 0.985)
(220.746, 0.985)
(222.140, 0.985)
(223.535, 0.986)
(224.929, 0.986)
(226.323, 0.986)
(227.717, 0.986)
(229.112, 0.986)
(230.506, 0.986)
(231.900, 0.986)
(233.294, 0.986)
(234.689, 0.986)
(236.083, 0.986)
(237.477, 0.986)
(238.871, 0.986)
(240.266, 0.986)
(241.660, 0.986)
(243.054, 0.986)
(244.448, 0.986)
(245.843, 0.986)
(247.237, 0.986)
(248.631, 0.986)
(250.025, 0.986)
(251.420, 0.986)
(252.814, 0.986)
(254.208, 0.986)
(255.602, 0.986)
(256.997, 0.986)
(258.391, 0.986)
(259.785, 0.986)
(261.179, 0.986)
(262.574, 0.986)
(263.968, 0.986)
(265.362, 0.986)
(266.756, 0.986)
(268.151, 0.986)
(269.545, 0.987)
(270.939, 0.987)
(272.333, 0.987)
(273.728, 0.987)
};

\legend{Non-tracking, Tracking}

\end{axis}
\end{tikzpicture}
        \subcaption{Number of storage access}
        \label{fig:StorageAccess}
    \end{minipage}%
    
     \vspace{0.1in}
    \caption{CDF highlights \tool features: (a) successor functions, and (b) storage access, in tracking vs non-tracking contexts.}
    \label{fig:influentialfeatures}
    \vspace{-0.1in}
\end{figure}
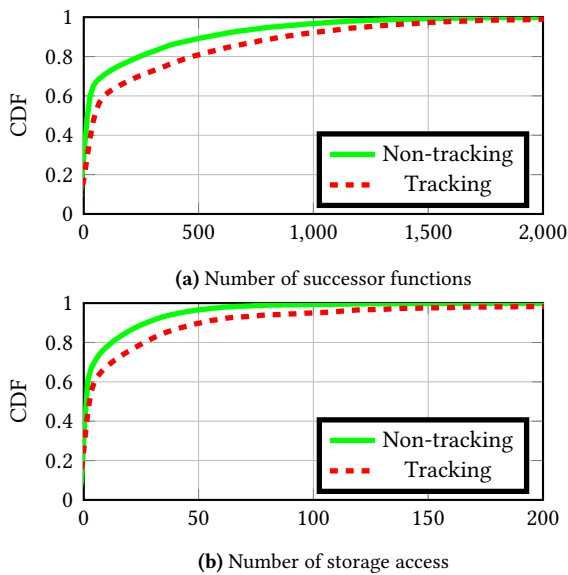

To better understand which features are potentially useful to \tool, we analyze feature importance using information gain (see Table ~\ref{table:information-gain}). 
The most important feature is the number of successor functions, which is a structural feature, followed by the number of local storage accesses, which is a contextual feature.
The number of successor functions provides insights into the calling context of the function.
Successor functions show how connected a function is to other parts of the code. 
Figure \ref{fig:NumOfMethSucces} shows that tracking JS functions tend to have more successor functions than non-tracking functions.
The average number of successor functions is 252 for tracking functions and 148 for non-tracking functions.
The higher number of successor functions in tracking activities indicates more complex behavior, like gathering, processing, and transmitting personal data.
Additionally, the number of storage accesses is the second most important feature in distinguishing tracking functions from non-tracking. 
Previous research recognizes that storage APIs are commonly employed by trackers \cite{Siby22WebGraph}.
Figure \ref{fig:StorageAccess} shows that tracking JS functions have a higher frequency of storage accesses than non-tracking functions.
The average storage access is 15 in tracking functions and 8 in non-tracking functions. 

\begin{table}[t]
\small
    \centering
    \scalebox{0.95}{\begin{tabular}{ |c|c| c | c| c| c|}
\hline
\small\textbf{Initiator} & \small\textbf{Non-initiator} & \small\textbf{Context} & \small\textbf{Precision} & \small\textbf{Recall} & \small\textbf{F1} \\
\small\textbf{Functions} & \small\textbf{Functions} & \small{} & \small{} & \small{} & \small\textbf{Score} \\
\hline
   \ding{52} & \ding{55} & \ding{55} & 68.2\% & 65.1\% & 66.9\% \\
   \cline{1-3}
   \ding{52} & \ding{52} & \ding{55} & 55.8\% & 99.7\% & 71.5\% \\
    \cline{1-3}
   \ding{52} & \ding{52} & \ding{52} & 94.3\% & 98.0\% & 96.2\% \\
\hline
\end{tabular}

}
    \vspace{0.15in}
    \caption{Ablation analysis results of \tool's features in terms of precision, recall, and F1-score.}
    \label{table:ablation}
    \vspace{-0.30in}
\end{table}
\noindent\textbf{Ablation analysis.}
Next, we evaluate the impact of various graph configurations and features.
We summarize our findings in Table \ref{table:ablation}. 
First, we find that solely incorporating initiator functions in the graph while excluding non-initiator functions lowers the F1-score of \tool by 29.3\%.  
This reduction is primarily due to the limitation that focusing solely on initiator functions omits the broader context essential to distinguish between tracking and non-tracking activities.
Next, we include both initiator and non-initiator functions but exclude the execution context features.
%
Under this configuration, although precision decreases, there is a notable increase in the recall. 
This shows that the model now possesses a broader code coverage — it is identifying a greater number of tracking functions, thereby elevating recall. 
However, this leads to an increase in false positives. 
%
%
We observe that a call to the same function in tracking and non-tracking contexts has the same representation in the graph; thus, the extracted features are not distinctive enough to accurately segregate tracking functions from non-tracking ones.
Finally, we include all functions and contextual features.
%
This setting surpasses the performance of all prior configurations owing to the enhanced code coverage and execution context.
The integration of calling context in \tool results in the best precision and overall F1 score.
This underscores the pivotal role of JS dynamic execution context with both initiator and non-initiator functions, emphasizing its importance in precisely identifying tracking JS functions.

\subsection{Robustness}
We evaluate the robustness of \tool in detecting tracking functions amid manipulation attempts such as JS  code obfuscation and code modifications, as well as in the context of enhanced code coverage.
\noindent\textbf{Code obfuscation.}
JS obfuscation is often employed to conceal the meaning of the code, making it harder to decipher for those who may try to reverse engineer or modify it \cite{xu2012power, choi2009automatic}. 
%
%
From our dataset of 10K websites, we randomly selected 10\%, which includes 15,939 unique scripts.
We then obfuscate these scripts using the {\tt obfuscater.io} \cite{obfuscator}, with the configuration shown in Listing \ref{lst:obfuscation}.

JS obfuscation poses several challenges for \tool.
First, it modifies the names of JS functions using hexadecimal notation, making it difficult for \tool to identify and track the execution of these functions accurately.
Second, it alters the JS execution call stack through the use of multiple techniques, such as control flow flattening \cite{wang2014technique,balachandran2016control} and self-defending \cite{ren2023empirical, qin2023transast}. 
Control flow flattening involves breaking down JS functions into smaller basic blocks and then rearranging these blocks to change the order of execution. 
This can make it difficult for \tool to understand the flow of the code and track the execution of different functions.
Self-defending adds protection code to the program that can detect if it is being debugged or modified and attempt to avoid execution in these cases. 
The protection code can prevent \tool from gathering the necessary information by changing the execution call stack of JS activity.
\begin{lstlisting}[language=Java, label={lst:obfuscation}, caption=Configuration used to obfuscate the scripts using obfuscator.io.]
compact: true,
controlFlowFlattening: true,
controlFlowFlatteningThreshold: 1,
deadCodeInjection: true,
deadCodeInjectionThreshold: 1,
disableConsoleOutput: true,
identifierNamesGenerator: 'hexadecimal',
rotateStringArray: true,
selfDefending: true,
stringArray: true,
stringArrayThreshold: 1,
transformObjectKeys: true
\end{lstlisting}

\begin{figure*}[!t]
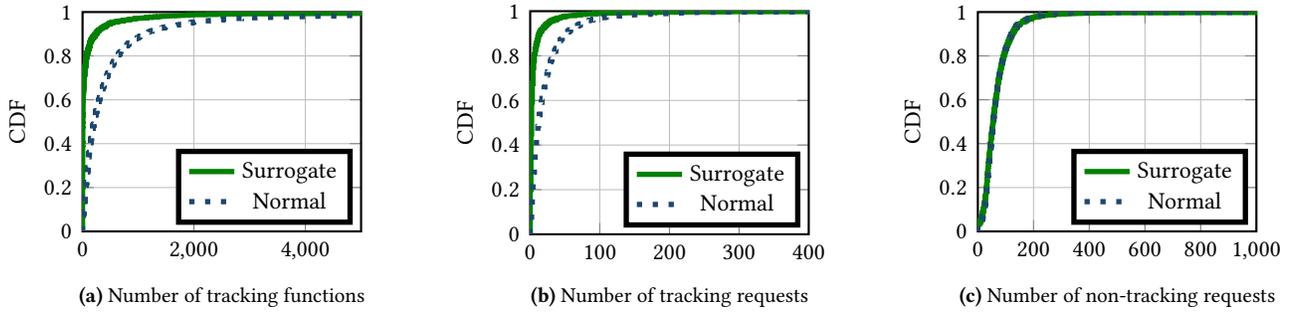

        \begin{minipage}[t]{0.33\textwidth}
        \input{plots/surrogate-tracking-functions}
        \subcaption{Number of tracking functions}
        \label{fig:functions}
    \end{minipage}
    \begin{minipage}[t]{0.33\textwidth}
        \input{plots/surrogate-tracking-requests}
        \subcaption{Number of tracking requests}
        \label{fig:tracking}
    \end{minipage}
    \begin{minipage}[t]{0.33\textwidth}
        \input{plots/surrogate-functional-requests}
        \subcaption{Number of non-tracking requests}
        \label{fig:non-tracking}
    \end{minipage}%
    \vspace{0.15in}
    \caption{CDF illustrates pre- and post-surrogate replacement metrics for websites: tracking functions, tracking requests, non-tracking requests.}
    \label{fig:surrogate-replacement}
    \vspace{-0.02in}
\end{figure*}
Table \ref{table:robustness} presents the classification results of \tool on the obfuscated data using the aforementioned configuration.
Overall, obfuscating the JS code reduces the precision by only 0.8\% and recall by 7.6\%, leading to a 4.3\% reduction in the overall F1 score.
The marginal drop in recall can be attributed to the model's increased likelihood of missing additional tracking functions.
This can be traced back to the non-operational ``garbage" functions in the stack, which the model finds challenging to accurately identify.
%
Given \tool's reliance on the dynamic execution context over static features, its precision remains unaffected by JS obfuscation.
\tool's features are robust against JS obfuscation, with the top-5 features listed in Table \ref{table:vulnerable} showing minimal change, averaging only 2.92\% change.
For instance, the average number of successor functions for the tracking function is 252.1 without obfuscation and 257.5 with obfuscation, representing a percentage difference of 2.14\%.
Similarly, the average number of storage accesses for the tracking function is 15.6 without obfuscation and 16.2 with obfuscation, representing a percentage difference of 3.84\%.
To further enhance \tool's robustness again script obfuscation, future work could incorporate adversarial training by explicitly adding obfuscated scripts in the training set \cite{devine2021adversarial}.

\noindent\textbf{Enhanced code coverage.}
To assess the impact of enhanced code coverage on \tool's performance, we conduct a more exhaustive crawl on a randomly selected subset, making up 10\% of our original 10K website corpus.
During this analysis, in addition to assessing the landing page, we explored five distinct internal pages of each website. 
We implement bot mitigation strategies that include simulating mouse movements at five unique offsets using the {\tt move\_by\_offset(x, y)} function.
Furthermore, we incrementally scroll through the webpage using the {\tt window.scrollBy()} function. 
Upon completing the website crawl, we utilize the \tool to classify the involved JS functions. 
As summarized in Table \ref{table:robustness}, our results indicate a 5.9\% decline in precision, which led to a 4.2\% reduction in the F1-score.
Consequently, the data reveals only a minor decrease in both precision and the overall F1-score for \tool, suggesting that it maintains a reliable level of accuracy even with expanded coverage.

\begin{table}[t]
\small
    \centering
    \scalebox{0.98}{\begin{tabular}{| l |r r |r|}
\hline
\textbf{Feature}& \textbf{Normal} &\textbf{Obfuscation} &\textbf{$\delta$}\\
& \textbf{Mean} &\textbf{Mean}&\textbf{Mean}\\
\hline
Number of successor functions & 252.14 & 257.46 & 5.32 \\
Number of local storage accesses & 15.62  & 16.28 & 0.66\\
Number of descendant nodes & 191.83 & 185.02 & 6.81\\
Number of in-edges & 2.02 & 2.03 & 0.01\\
Number of web API calls & 5.47 & 7.31 & 1.84\\
\hline

\end{tabular}}
    \vspace{0.2in}
    \caption{The change in the mean of the top five most important features with JS obfuscation ($\delta$).}
    \label{table:vulnerable}
    \vspace{-0.30in}
\end{table}

\subsection{Comparison with Existing Countermeasures}
We compare the performance of \tool against two state-of-the-art baselines: \webgr and SugarCoat, evaluating their precision and recall specifically for identifying tracking functions. 
Later, in section \ref{subsec: surrogate generation},  we assess the impact on website functionality, with a focus on website breakage.

\noindent\textbf{\webgr comparison.} 
\webgr, by default, classifies tracking script URLs, presuming all functions in the script have the same label as the script's URL.
In contrast, \tool uniquely classifies each JS function individually.
The comparison between \webgr and \tool is conducted on the same dataset.
We crawl each website once and then build two different graphs – one for \webgr \footnote{We use Structural + Flow features for \webgr since the Content features are not robust, as shown in the \webgr paper.} and one for \tool, as the graph representations differ between the two approaches.
The appendix \ref{sec:webgraph-graph} provides a detailed account of the key differences in graph representations between \webgr and \tool.
Table \ref{table:robustness} provides a summary of the results for \tool and \webgr. 
%
\tool outperforms \webgr, which classifies script-URL, in classifying tracking JS functions within mixed scripts with 45.0\% higher precision, 31.6\% higher recall, and an overall 39.7\% better F1-score.
\webgr is unable to fully capture the communication edges between the scripts present in the call stack during a tracking activity. 
As a result, the dynamic execution context surrounding data transfer and the initiator script, which sends the data to the server, is not captured and modeled.
Furthermore, tracking at the script-level granularity obscures important features that are associated with fine-grained JS functions, making it difficult to distinguish tracking and non-tracking activity in mixed scripts~\cite{amjad2023blocking}.
Although \webgr claims a 92\% accuracy rate in its paper, this is largely because it is tailored to identify tracking script URLs. 
Its limitations become evident when dealing with mixed scripts, underscoring the need for function-level granularity.  
\noindent\textbf{SugarCoat comparison.} 
SugarCoat follows a three-step process. 
First, a developer curates a list of mixed scripts containing both tracking and non-tracking functions. 
Second, SugarCoat generates a behavioral graph for these target mixed scripts, pinpointing six privacy-relevant API accesses to specific locations within the JS source code by analyzing the collected data. 
Third, SugarCoat creates a surrogate of the target mixed scripts, redirecting the identified API accesses to manually crafted mock implementations \cite{SugarCoatmocks}.
Overall, SugarCoat requires significant manual effort, including the identification of mixed scripts and crafting handwritten mock API implementations.
In contrast, \tool relies on a detection model to automatically identify mixed scripts and does not rely on handwritten mock API implementations.
Thus, it is not surprising that SugarCoat has approximately two hundred surrogate scripts \cite{SugarCoatresources}.

Since SugarCoat lacks an automated detection model, for the sake of comparison, we assume that the functions modified in its publicly available surrogates \cite{SugarCoatresources} are classified as tracking functions. 
For example, in the SugarCoat repository, we find a script \url{ads.google.com/aw/JsPrefetch?origin=lead_in_page} \cite{SugarCoat-example}. 
Its function {\tt prefetchJS()} is replacing the default {\tt localStorage} API with the mock implementation, indicating that SugarCoat classified this function as tracking.
In contrast, \tool utilizes its classification model for this purpose.
Table \ref{table:robustness} provides a summary of the results for \tool and SugarCoat. 
\tool outperforms SugarCoat in detecting tracking JS functions with 71.3\% higher precision, 75.8\% higher recall, and an overall 73.4\% better F1-score.
As SugarCoat is designed to address only mixed scripts, we precisely compute the precision and recall for only mixed scripts.
Even in this favorable comparison scenario, SugarCoat achieves 27.5\% precision, 35.1\% recall, and an overall 30.9\% F1-score in detecting tracking JS functions.
This clearly shows that SugarCoat's manual detection and dependency on six manually crafted mock implementations for tracking functions makes it infeasible to apply at scale.

\subsection{Surrogate Generation and Replacement}
\label{subsec: surrogate generation}
We assess the feasibility of \tool's real-time deployment through the evaluation of its surrogate generation and replacement strategy.
For our evaluation, surrogates are generated for a randomly selected 50\% of the 10K websites, after which we re-visit each site, substituting the original script with the surrogate prepared by \tool.
We utilize the surrogate replacement strategy as outlined in Section \ref{sec:methodology}, implemented as a Chrome extension.
%


\noindent\textbf{Surrogate generation.}
As elaborated in Section \ref{sec:methodology}, 
\tool locates the tracking function call within the script source code and substitutes it with a mock function call, effectively neutralizing it.
During the surrogate generation, we calculate four parameters to assess the efficacy of the technique.
First, we measure the average number of tracking function calls per webpage that can be successfully neutralized, which is 122.3.
Second, we measure the average number of tracking function calls per webpage that cannot be neutralized due to the limitations of our instrumentation,  which fails to capture the script source code. The average number of such tracking functions per webpage is 3.1.
Third, we measure the average number of tracking function calls per webpage that cannot be neutralized due to script inlining—\ie embedded within the HTML of the page. The average number of such tracking functions per webpage is 4.7.
Finally, we measure the average number of tracking function calls per webpage that cannot be neutralized owing to the dynamism in the JS \cite{wei2014state, wei2016empirical}, rendering us unable to confirm the line number and column number at run-time. The average number of such tracking functions per webpage is 14.8.
In summary, our surrogate generation technique successfully neutralizes an average of 84.4\% of classified tracking functions per webpage.

\noindent\textbf{Time and space analysis.} The average time required to generate a surrogate is 1.31 seconds. 
The size of an average surrogate generated by \tool is 164 KB. 
For 50\% of the 10K websites, a total space of 9.6 GB is required.
However, users of \tool do not need to download all surrogate scripts at once. 
We discuss in more detail in Section \ref{sec:limitations} how \tool can be implemented more efficiently. 
For example, it can cache surrogate scripts for commonly visited websites and potentially pre-fetching popular scripts.

\noindent\textbf{Surrogate replacement.}
After generating the surrogates, we evaluate both privacy and usability metrics across 50\% of these websites, both before and after surrogate deployment. 
We calculate the average number of tracking functions, as well as tracking and non-tracking requests per website. 
Furthermore, we carry out a performance and user-centered manual breakage analysis to verify the usability of websites after surrogate deployment.
Originally, the average number of tracking functions per website is 153.6,  
which drops to 38.5 after deploying the surrogates, representing an 80.1\% reduction per website. 
Figure \ref{fig:functions} displays the CDF that shows a significant decrease in tracking functions 
after surrogate deployment.
Moreover, the average count of tracking requests per website is initially 28.0. This decreases to 8.4 post-deployment, representing a 76.9\% reduction per website.
Figure \ref{fig:tracking} displays the CDF that shows a significant decrease in tracking requests on the majority of websites following surrogate deployment. 
Finally, the average number of non-tracking requests per website is 74.3 and drops minimally to 71.5 after deploying the surrogates. 
Figure \ref{fig:non-tracking} displays the CDF, illustrating the number of non-tracking requests against the website distribution. 
The overlapping lines in the graph indicate negligible impact on website usability.

\noindent\textbf{Performance analysis.}
We evaluate the performance overhead across 50\% of these websites, both before and after surrogate deployment, utilizing Selenium \cite{selenium} to extract standard page performance metrics for each visit. 
These metrics include JS memory usage and the timing of key page load events.
Table \ref{table:performance} summarizes the results. 
Regarding JS memory usage, we observe a decrease of 32\% in total and 25\% in used JS heap memory. 
The total JS heap size represents the entire memory allocation for JS execution, while the used portion reflects the memory actively utilized by JS on the webpage.
This decrease is primarily due to the neutralization of tracking function calls, which return an empty response instead of executing the original memory-consuming function.

A standard benchmark ~\cite{UserExp} for user-perceived web page performance measures the timeline of key events marking various stages in the browser's page load and rendering process. With \tool's surrogates, these events are completed on average 32 milliseconds earlier.
These improvements are seen across several metrics: DOM content load time, which indicates the time from page load start to complete, HTML parsing, and initial DOM interactive time.
DOM interactive time indicates the duration until the DOM is fully prepared for user interaction, and load event time indicates the total time to completely load all resources, such as images and CSS, signifying the full usability of a webpage.
This overall improvement is expected, as our neutralized tracking function calls avoid invoking the original tracking functions, performing less work by simply returning an empty response.

\noindent\textbf{User-centric breakage analysis.}
We conduct a qualitative manual analysis of \tool, following methodologies from previous studies ~\cite{amjad2023blocking, munir2023cookiegraph}.
We select 50 webpages from the top-10K websites, specifically those hosting scripts mentioned in the exception rules of filter lists\footnote{The EasyList project tags git commit messages addressing compatibility fixes with “P:” - see  \url{https://github.com/easylist/easylist/commits.}} and addressed with SugarCoat's six mock API implementations \cite{SugarCoatmocks}.
These are mainly mixed scripts; content blockers typically avoid blocking them to prevent website breakage, even though allowing them may compromise user privacy.

One of the authors evaluates all fifty webpages.
We recruit ten independent evaluators for breakage assessments.
Each independent evaluator evaluates five distinct websites from our sample.
%
Each webpage is evaluated by at least two different evaluators.
\begin{table}[t]
\small
    \centering
    \scalebox{0.9}{\begin{tabular}{ |c|r| r|}
\hline

\multirow{2}{*}{\textbf{Metrics}}& \textbf{Normal} & \textbf{Surrogate} \\
    & \textbf{(Mean, Median)} & \textbf{(Mean, Median)} \\
\hline
    \textbf{\textit{JavaScript Memory Usage}}&&\\
    Heap Total Size& 18.79 \textit{MB}, 13.92 \textit{MB} & 12.86 \textit{MB}, 9.18 \textit{MB}\\
    Heap Used Size& 12.95 \textit{MB}, 10.67 \textit{MB} & 9.74 \textit{MB}, 7.39 \textit{MB}\\
    \hline
    \textbf{\textit{Performance Event Timing}}&&\\
    DOM Content Loaded& 1,402 \textit{ms}, 1,136 \textit{ms} & 1,290 \textit{ms}, 1,112 \textit{ms}\\
    DOM Interactive& 1,067 \textit{ms}, 931 \textit{ms} & 1,162 \textit{ms}, 1,005 \textit{ms}\\
    Load Event& 2,641 \textit{ms}, 1,888 \textit{ms} & 2,562 \textit{ms}, 1,804 \textit{ms}\\

\hline

\end{tabular}}
    \vspace{0.2in}
    \caption{Performance analysis for the post-surrogate replacement.}
    \label{table:performance}
    \vspace{-0.2in}
\end{table}

The webpages are evaluated in four different configurations: (1) Control, which displays the default webpage without any blocking; (2) \webgr, highlighting the limitations of advanced script-level blocking in mixed script scenarios; (3) SugarCoat, highlighting its effectiveness, though it faces scalability challenges, especially with limited mock API implementations \cite{SugarCoatmocks} (4) \tool, showing a reduced impact on website functionality compared to \webgr, while being on par with SugarCoat in terms of handling mixed scripts.
We ensure high coverage for \tool by ensuring that 98\% of the surrogate scripts are triggered at page load.
This ultimately highlights that \tool significantly advances the handling of mixed scripts compared to current state-of-the-art approaches.
We present \webgr, SugarCoat, and \tool in a randomized sequence to each evaluator, revealing only the control configuration.

We ask evaluators to classify breakage into four categories: navigation (moving between pages), SSO (initiating and maintaining login state), appearance (visual consistency), and miscellaneous (such as chats, search, and shopping cart).
Each evaluator labels breakage as either major or minor for each category: Minor breakage occurs when it is difficult but not impossible for the evaluator to use the functionality. 
Major breakage occurs when it is impossible to use the functionality on a webpage.
The inter-evaluator agreement was 95.5\%, suggesting substantial agreement.
Conflicts are resolved by one of the authors revisiting those websites.
Conflicts mainly stem from ambiguity between appearance and miscellaneous categories, such as the interchangeable reporting of a missing ad overlay in both.

\begin{figure}[t]
    \centering
        \includegraphics[width=0.49 \textwidth]{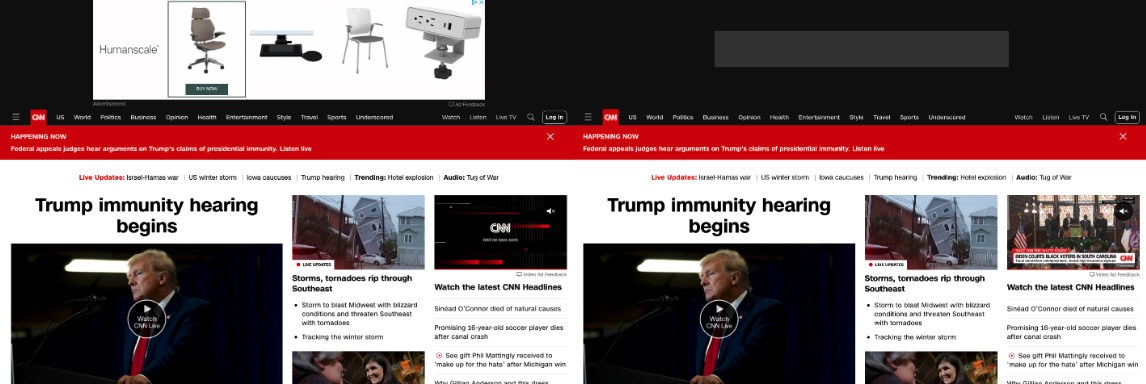} 
         \begin{tikzpicture}[overlay]
          \draw[red,thick,rounded corners] (-4.1,2.5) rectangle (0,3.2);
          \draw[red,thick,rounded corners] (0.3,2.5) rectangle (4.3,3.2);
        \end{tikzpicture}
        \caption{In the appearance category, 4\% reported minor breakage attributed to \tool is, in fact, not a breakage but simply a leftover overlap after tracking is removed.}
        \label{fig:minor_notjs}
\end{figure}
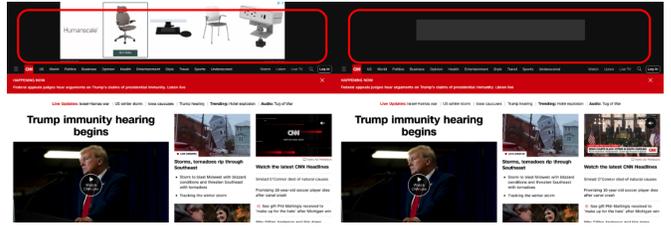

Table \ref{table:breakageanalysis} summarizes the results in each category.
\webgr causes major breakage on 6\% and minor breakage on 10\% of the webpages.
In contrast, \tool, only causes minor breakage on 8\% of the webpages, without any major breakage.
As compared to \webgr, \tool significantly improves the major breakage in the navigation category.
For example, on the webpage {\tt bbc.com}, the search bar in the navigation menu disappears with \webgr, resulting in major breakage. 
This major breakage stems from the blocking of the mixed script hosted by {\tt bbc.com}. \tool effectively mitigates tracking in the mixed script by generating a surrogate replacement. 
While \tool performs comparably to SugarCoat in terms of major breakage, it results in 4\% more minor breakage in the appearance category. 
As shown in Figure \ref{fig:minor_notjs}, this minor breakage is due to a missing ad overlay left after successfully neutralizing the tracking function. 
Overall, \tool matches SugarCoat in terms of breakage while being more scalable due to its automated surrogate generation as compared to SugarCoat's reliance on just six manually crafted mock implementations.


\begin{table}[t]
\small
    \centering
    \scalebox{0.9}{\begin{tabular}{ | l | c c  c c  r r| }

\hline
\textbf{Category}& \multicolumn{2}{c}{\textbf{\webgr}} & \multicolumn{2}{c}{\textbf{SugarCoat}} & \multicolumn{2}{c|}{\textbf{\tool}}\\
& Minor & Major & Minor & Major & Minor & Major\\
\hline
Navigation & \colorbox{green}{ 0\% }& \colorbox{red}{ 6\% }& \colorbox{green}{ 0\% }& \colorbox{green}{ 0\% } & \colorbox{green}{ 0\% }& \colorbox{green}{ 0\% } \\
SSO & \colorbox{yellow}{ 2\% }& \colorbox{red}{ 2\% }& \colorbox{yellow}{ 2\% }& \colorbox{green}{ 0\% }& \colorbox{yellow}{ 2\% }& \colorbox{green}{ 0\% }\\
Appearance & \colorbox{yellow}{ 4\% }& \colorbox{green}{ 0\% }& \colorbox{green}{ 0\% }& \colorbox{green}{ 0\% }& \colorbox{yellow}{ 4\% }& \colorbox{green}{ 0\% }\\
Miscellaneous & \colorbox{yellow}{ 4\% }& \colorbox{red}{ 4\% }& \colorbox{yellow}{ 4\% }& \colorbox{green}{ 0\% }& \colorbox{yellow}{ 4\% }& \colorbox{green}{ 0\% }\\

\hline
\end{tabular}}
    \vspace{0.2in}
    \caption{Qualitative manual analysis for 50 webpages using \tool, SugarCoat, and \webgr, showing \% of \colorbox{green}{No}, \colorbox{yellow}{Minor}, and \colorbox{red}{Major} breakages in navigation, SSO, appearance, and miscellaneous categories. \textbf{4\% of minor appearance breakages with \tool are due to missing ad overlay, as shown in Figure \ref{fig:minor_notjs} in the appendix}.
        }
    \label{table:breakageanalysis}
    \vspace{-0.3in}
\end{table}


\section{In the Wild Deployment}
\label{sec:deployment}
In this section, we analyze the tracking JS functions detected by \tool on top-10K websites.

\noindent\textbf{Prevalence of tracking functions.}
\tool classifies 32.1\% of the 2,088K JS functions in our dataset as tracking. 
We find that these tracking functions are present in the scripts served by 8,587 unique domains.
Among these tracking functions, 8.2\% are anonymous functions, 18.3\% are part of inline scripts, and 1.5\% are part of eval scripts.
Table \ref{table:deployment} in the appendix lists the top-5 most prevalent tracking functions across top-10k websites.
For instance, the function {\tt "Z.D"} from the script {\tt analytics.js} hosted by {\tt google-analytics\\.com} appears on 56\% of the websites. 
This function, on average, invokes cookie setter 3.3 times and cookie getter 20 times.
Similarly, function {\tt "c"} from the script {\tt fbevents.js}, hosted by {\tt connect.face\\book.net}, appears on 21.5\% of the websites. 
Its typical calling context contains 3.9 closures and  6.6 get attribute ({\tt getAttribute}) calls, along with 2.5 cookie accesses.

\noindent\textbf{Characteristics of tracking functions in mixed scripts.}
%
We find that 13.4\% of all scripts are mixed, aligning with previous studies \cite{Amjad2021TrackerSift, amjad2023blocking}, while 62.3\% of websites incorporate at least one mixed script. 
On average, a website contains around 2.42 mixed scripts.
The overview of the domains (eTLD + 1) serving the most mixed scripts is provided in the appendix (Table \ref{table:mixedscript}).
Notably, 70.6\% of the mixed scripts are served from third-party domains\footnote{The domain of the script's URL differs from the top-level URL of the page.}.

Our  analysis of the top-100 mixed scripts reveals that they are commonly included in the first-party context, enabling the setting of ghost first-party cookies \cite{Sanchez2021JourneyToCookies}.
%
These scripts set 14,867 {\tt ghost} first-party cookies, out of which 150 are found to be tracking after running CookieGraph \cite{munir2023cookiegraph}, a tool designed to detect first-party tracking cookies.
\tool classified 83\% of JS functions in these mixed scripts as tracking functions that are either setting or getting these ghost first-party tracking cookies \cite{munir2023cookiegraph}.
For example, \tool detected the tracking function {\tt "a"} within the script {\tt launch-*.min.js} which is accessing the ghost first-party tracking cookie {\tt mbox} on  {\tt adobe.com}.
As another example, \tool detected the tracking function {\tt "o"} within the script {\tt opus.js} which is accessing the ghost first-party tracking cookie {\tt A1S} on {\tt yahoo.com}.

To illustrate how \tool tackles our threat model, consider these two types of mixed scripts.
The script named {\tt webpack-*.js} is served by the domain {\tt cloudfront.net}. 
\tool detects tracking and non-tracking functions within this mixed script. 
Specifically, the function {\tt "t"} accesses local storage four times, sets it six times, and attaches event listeners to 244 different DOM elements, detected as tracking.
In contrast, the {\tt "a"} function in the same script avoids interactions with both local storage and the DOM, detected as non-tracking. 
Another scenario involves the script {\tt app.js} from the domain {\tt acsbapp.com}, which contains the function {\tt "\_e"}. 
The behavior of this function depends on its dynamic execution context, detected by \tool. 
In a non-tracking calling context, the number of closures and local variables for the function are 3 and 1, respectively.
However, in a tracking calling context, these values are 0, emphasizing the importance of context to distinguish between tracking and non-tracking behaviors.

\section{Discussion}
\label{sec:limitations}
In this section, we discuss some opportunities for future work and limitations of \tool.

\noindent\textbf{User interaction limitations.} The interactions captured by \tool depend on the diversity and intensity of user activity, such as scrolling or clicking on internal pages.
\tool may miss certain tracking functions due to limited user interactions. 
To mitigate this, we propose to use forced execution \cite{kim2017j, peng2014x, tang2018dual} in the future to improve the completeness of the webpage's graph.

\noindent\textbf{Browser-specific deployment.} \tool uses the Chrome browser and Chrome-based extensions to collect data due to its popularity. 
Extensions on other browsers (Firefox \cite{Firefox}, Safari \cite{Safari}, Edge \cite{Microsoft-Edge}) have different permissions and access to varying sets of information about a webpage's activity. 
Porting \tool on other browsers may require additional engineering. 

\noindent\textbf{Expanding beyond main thread execution.} \tool presently captures only the dynamic execution context of the main thread, leaving out service workers that operate in a separate execution context. 
Enhancing \tool to include these workers is a targeted area for future work, which will extend \tool's ability to capture additional webpage's activity.

\noindent\textbf{Resource-constrained environments.} While \tool is adept at generating surrogate scripts at scale, its deployment on devices with limited storage presents a challenge. 
We can implement specific strategies for devices with limited storage.
First, \tool can prioritize caching popular scripts that are frequently accessed, ensuring these replacements are readily available locally to reduce reliance on server requests.
Second, we can employ selective prefetching techniques to allow \tool to anticipate and fetch surrogate scripts for popular websites or commonly visited domains in advance. 
Third, for less frequently accessed or uncached scripts, \tool can dynamically request them.

\noindent\textbf{Mixed function analysis.} In our ground-truth only 3.9\% of the functions are mixed, \ie involved in both tracking and non-tracking activities. 
Among these, a mere 0.8\% are integral in contexts requiring the blocking of tracking activity, while the remainder can be left unblocked by targeting other tracking functions in the activity's execution chain.
Future work can focus on a more fine-grained analysis of these mixed functions, examining individual statements within the functions.

\noindent\textbf{Usage of bundlers.} 
To investigate mixing tracking with non-tracking functions, we identify bundled scripts within mixed scripts using a high-precision heuristic from the Web Almanac~\cite{JavaScript2022WebAlmanac}, focusing on the {\tt webpackJsonp} keyword in scripts\footnote{\url{https://v4.webpack.js.org/configuration/output/\#outputjsonpfunction}}. This keyword, default in {\tt JSONP} functions, indicates asynchronous loading of bundled scripts, commonly with Webpack \cite{webpack}. We observe that 20.2\% of mixed scripts are Webpack-bundled, comparable to functional scripts (20\%) but exceeding tracking scripts (10\%). However, potential imperfect recall exists as {\tt webpackJsonp} can be altered in Webpack's settings, and this heuristic doesn't cover scripts bundled via tools like Parcel \cite{Parcel}, Rollup \cite{Rollup}, or Browserify \cite{Browserify}, thus underrepresenting the actual count of mixed, bundled scripts. Future investigations could enhance script-mixing understanding by developing more precise heuristics for bundled script identification.

%

%

\section{Conclusion}
\label{sec:conclusion}
%
\tool advances the state-of-the-art  by identifying tracking JS functions in mixed scripts and generating surrogates to replace mixed scripts. 
In doing so, \tool achieves 94\% precision and 98\% recall, surpassing state-of-the-art in terms of accuracy, robustness, and minimizing breakage. 
%
%
Our analysis shows that \tool can detect and neutralize mixed scripts that are not blocked by filter lists due to breakage concerns despite being known to engage in tracking activities.
As the arms race evolves, finer-grained tracker blocking tools such as \tool will be imperative to handle mixed scripts.



%

\bibliographystyle{ACM-Reference-Format}
\bibliography{bib}

\section{Appendix}
\label{sec:appendix}

\subsection{Shortcomings in \webgr's graph representation.}\label{sec:webgraph-graph}
\begin{figure}[h]
    \centering
      \includegraphics[width=0.43\textwidth]{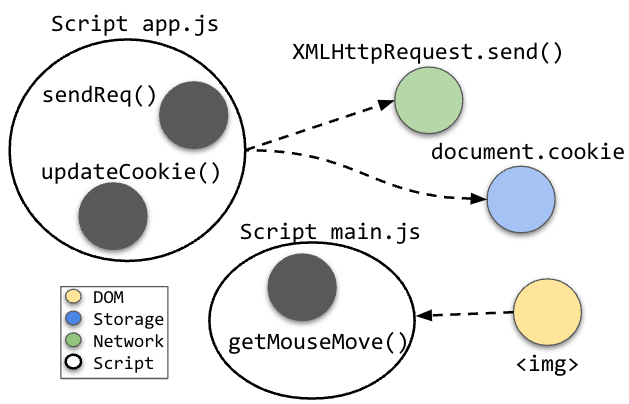}
        \caption{\webgr's graph representation of tracking mouse movement.}
        \label{fig:scenerio-webgraph}
        \vspace{0.1in}
\end{figure}

Figure \ref{fig:scenerio-webgraph} shows a graph constructed by \webgr for the execution flowchart in Figure \ref{fig:scenerio}. 
There are key differences in the graph representation of \webgr compared to \tool. 
First, nodes in the \webgr's graph representation are at the script-level instead of the function-level. 
Thus, in \webgr, all functions within a script will be assigned the same label and features as the script, leading to over-approximation. 
Second, \webgr does not capture the JS execution call stack, missing all the edges between caller-callee scripts, such as the edge between script {\tt main.js} and {\tt app.js}. 
Third, \webgr includes two storage APIs (\ie local storage and cookies) and does not include the rest of the Web APIs that are implemented in \tool. 
It also lacks the calling context of the function when it was called, such as the arguments of the function. 
As shown earlier, this calling context at finer granularity plays a crucial role in detecting tracking functions.
In short,  \webgr's graph representation misses an average of 262.3 nodes and 673.5 edges per webpage which are key to fine-grained detection of tracking JS code. 

\begin{table}[h]
\small
    \centering
    \scalebox{1}{\begin{tabular}{ |c |c |c|}
\hline
\multirow{2}{*}{\textbf{Actual}}& \multicolumn{2}{c|}{\textbf{Predicted}}\\
\cline{2-3}
  & 0 (non-tracking) & 1 (tracking)\\
\hline
0 (non-tracking) & 216,996 & 8,011 \\
1 (tracking) & 2,720 & 133,325 \\

\hline
\end{tabular}}
    \vspace{0.2in}
    \caption{The classification report and corresponding confusion matrix for the \tool.}
    \label{table:confusionmatrix}
\end{table}
\begin{table}[!t]
    \centering
    \scalebox{0.9}{\begin{tabular}{| l| r | r|}
\hline
\textbf{Mix script}& \textbf{number of}& \textbf{number of}\\
\textbf{domain}& \textbf{mix scripts}& \textbf{websites}\\
\hline
{\tt googleapis.com } & 232 & 620 \\
{\tt tiqcdn.com} & 179 & 133 \\
{\tt cloudfront.net}& 129 & 186\\
{\tt google.com}& 120 & 615\\
{\tt facebook.net}& 51 & 418\\
{\tt taboola.com}& 42 & 145\\
{\tt gstatic.com}& 40& 746\\
{\tt trustarc.com}& 31& 39\\
{\tt primis.tech}& 23 & 21\\
{\tt intercomcdn.com}& 19 & 183\\
\hline
\end{tabular}}
    \vspace{0.2in}
    \caption{Top ten domains hosting mixed scripts and their presence on a number of websites.}
    \label{table:mixedscript}
\end{table}

\begin{table}[t]
\small
    \centering
    \scalebox{1}{\begin{tabular}{ |l |l| }
\hline
\textbf{Features} & \textbf{Type}\\
\hline
Number of nodes & Structural\\
Number of edges& Structural\\
Number of nodes/edges ratio& Structural\\
Number of edges/nodes ratio& Structural\\
Number of in-edges& Structural\\
Number of out-edges& Structural\\
Number of in+out-edges& Structural\\
Average degree connectivity& Structural\\
Closeness centrality (in/out)& Structural\\ 
Eccentricity& Structural\\
Number of descendant nodes & Structural\\
Number of ascendants nodes & Structural\\
Number of successor functions & Structural\\
Number of predecessor functions & Structural\\
Number of descendants with storage access & Structural\\
Number of ascendants with storage access & Structural\\
Number of descendants with web API access & Structural\\
Number of ascendants with web API access & Structural\\
Number of caller functions& Structural\\
Number of callee functions& Structural\\
Number of requests sent& Contextual\\
Parent is an eval script& Contextual\\
Is gateway function& Contextual\\
Number of local storage  (getter/setter)& Contextual\\
Number of cookie  (getter/setter)& Contextual\\
Number of Web API  (getter/setter)& Contextual\\
Number of arguments& Contextual\\
Number of local variables& Contextual\\
Number of global variables& Contextual\\
Number of closure variables& Contextual\\
Number of addEventListener calls& Contextual\\
Number of removeEventListener calls& Contextual\\
Number of getAttribute calls& Contextual\\
Number of setAttribute calls& Contextual\\
Number of removeAttribute calls& Contextual\\
Number of getAttribute calls& Contextual\\

\hline
\end{tabular}}
    \vspace{0.2in}
    \caption{The complete list of features of \tool.}
    \label{table:Allfeatures}
\end{table}

\begin{table*}[!t]
    \centering
    \scalebox{0.9}{\begin{tabular}{| l l l c c c c c|}
\hline
\textbf{Tracking}& \textbf{Script}& \textbf{Org.}& \textbf{\% of} & \textbf{Average}& \multicolumn{3}{c|}{\textbf{Important Features (Mean, Median, Mode)}}\\
\textbf{Function}& \textbf{Domain}& &\textbf{Websites}& \textbf{ Reqs Sent}&&&\\
\hline
\multirow{2}{*}{\tt Z.D} & \multirow{2}{*}{google-analytics.com}& \multirow{2}{*}{Google} & \multirow{2}{*}{56.0\%} & \multirow{2}{*}{6.89} & cookie-setter & cookie-getter & addEventListener  \\
&&&&& (3.3, 3.0, 3) & (20.0, 18.0, 19)  & (0.49, 0, 0)   \\

\cline{1-8}
\multirow{2}{*}{\tt c} & \multirow{2}{*}{connect.facebook.net}& \multirow{2}{*}{Facebook} & \multirow{2}{*}{21.5\%} & \multirow{2}{*}{7.59} & cookie-getter & num-closure & getAttribute  \\
&&&&& (2.5, 2.0, 2) & (3.9, 4.0, 4)  & (6.6, 6.0, 6)   \\

\cline{1-8}
\multirow{2}{*}{\tt G} & \multirow{2}{*}{tpc.googlesyndication.com}& \multirow{2}{*}{Google} & \multirow{2}{*}{9.0\%} & \multirow{2}{*}{5.74} & addEventListener & num-closure & num-local  \\
&&&&& (0.9, 1.0, 1) & (1.8, 2.0, 2)  & (0.9, 1.0, 1)   \\

\cline{1-8}
\multirow{2}{*}{\tt x} & \multirow{2}{*}{snap.licdn.com}& \multirow{2}{*}{Microsoft} & \multirow{2}{*}{7.6\%} & \multirow{2}{*}{2.38} & cookie-setter & cookie-getter & num-closures  \\
&&&&& (1.0, 1.0, 1) & (1.1, 1.0, 1)  & (4.2, 5.0, 5)   \\

\cline{1-8}
\multirow{2}{*}{\tt UET} & \multirow{2}{*}{bat.bing.com}& \multirow{2}{*}{Microsoft} & \multirow{2}{*}{6.2\%} & \multirow{2}{*}{0} & storage-getter & cookie-getter & addEventListener  \\
&&&&& (1.1, 1.0, 1) & (1.3, 1.0, 1)  & (1.9, 2.0, 2)   \\

\hline
\end{tabular}}
    \vspace{0.2in}
    \caption{Top-five most prevalent tracking functions and their properties.}
    \label{table:deployment}
\end{table*}

\end{document}